\documentclass[fleqn,usenatbib]{mnras}

\usepackage{newtxtext,newtxmath}

\usepackage[T1]{fontenc}

\DeclareRobustCommand{\VAN}[3]{#2}
\let\VANthebibliography\thebibliography
\def\thebibliography{\DeclareRobustCommand{\VAN}[3]{##3}\VANthebibliography}

\usepackage{graphicx}	
\usepackage{amsmath}	
\usepackage{mathrsfs}
\usepackage{hyperref}
\usepackage{pdflscape}
\usepackage{float}
\usepackage{here}

\title[The hyper-Eddington AGN IRAS\,04416+1215]{The extreme properties of the nearby hyper-Eddington accreting Active Galactic Nucleus in IRAS\,04416+1215}
\author[A. Tortosa et al.]{
Alessia Tortosa $^{1}$\thanks{E-mail: alessia.tortosa@mail.udp.cl},
Claudio Ricci $^{1,2}$,
Francesco Tombesi $^{3,4,5,6}$,
Luis C. Ho $^{2,7}$,
Pu Du $^{8}$,
Kohei Inayoshi $^{2}$,
\newauthor
Jian-Min Wang $^{8,9,10}$,
Jinyi Shangguan$^{11,2}$,
Ruancun Li$^{2}$
\\
$^{1}$Núcleo de Astronomía de la Facultad de Ingeniería, Universidad Diego Portales, Av. Ejército Libertador 441, Santiago, Chile\\
$^{2}$Kavli Institute for Astronomy and Astrophysics, Peking University, Beijing 100871, China\\
$^{3}$Dipartimento di Fisica, Univerisità di Roma Tor Vergata, via della Ricerca Scientifica 1, I-00133 Roma, Italy\\
$^{4}$INAF – Osservatorio Astronomico di Roma, Via Frascati 33, 00040 Monte Porzio Catone, Italy\\
$^{5}$ Department of Astronomy, University of Maryland, College Park, MD 20742, USA\\
$^{6}$ NASA Goddard Space Flight Center, Greenbelt, MD 20771, USA\\
$^{7}$ Department of Astronomy, School of Physics, Peking University, Beijing 100871, China\\
$^{8}$ Key Laboratory for Particle Astrophysics, Institute of High Energy Physics, Chinese Academy of Sciences, 19B Yuquan Road, Beijing 100049, China\\
$^{9}$ School of Astronomy and Space Sciences, University of Chinese Academy of Sciences, 19A Yuquan road, Beijing 100049, China\\
$^{10}$ National Astronomical Observatory of China, 20A Datun Road, Beijing 100020, China\\
$^{11}$  Max-Planck Institute for Extraterrestrial Physics (MPE), Giessen- bachstr. 1, 85748 Garching, Germany
}

\date{Accepted XXX. Received YYY; in original form ZZZ}

\pubyear{2021}

\begin{document}
\label{firstpage}
\pagerange{\pageref{firstpage}--\pageref{lastpage}}
\maketitle

\begin{abstract}
The physical properties of the accretion flow and of the X-ray emitting plasma, in supermassive black holes accreting at extreme Eddington rates, are still very unclear. Here we present the analysis of simultaneous \textit{XMM-Newton} and \textit{NuSTAR} observations of the hyper-Eddington Seyfert\,1 galaxy IRAS\,04416+1215, carried out in 2020. The main goal of these observations is to investigate the properties of the X-ray corona, as well as the structure of the accretion flow and of the circumnuclear environment, in this regime of extreme accretion. IRAS 04416+1215 has one of the highest Eddington ratio ($\lambda_{\rm Edd}\simeq 472$) in the local Universe. It shows an interesting spectral shape, very similar to the standard Narrow Line Seyfert 1 galaxy's spectra, with the presence of multi-phase absorption structure composed of three phases, whose estimate of the minimum and maximum distances suggests two different interpretations, one consistent with the three X-ray winds being co-spatial, and possibly driven by magnetohydrodynamical processes, the other consistent with the multi-phase winds being also multi-scale. The X-ray spectrum of IRAS\,04416+1215 also has a prominent soft excess component and a hard X-ray emission dominated by a reflection component. Moreover, our detailed spectral analysis shows that IRAS\,04416+1215 has the lowest coronal temperature measured so far by \textit{NuSTAR} ($kT_e=3-22$\,keV, depending on the model). This is consistent with a hybrid coronal plasma, in which the primary continuum emission is driven by pair production due to high-energy tail of the energy distribution of non-thermal electrons.\\ 
\end{abstract}

\begin{keywords}
keyword1 -- keyword2 -- keyword3
\end{keywords}



\section{Introduction}
Super Massive Black Holes (SMBH, M$_{\rm BH } > 10^5 $M$_{\odot}$) are found ubiquitously at the center of massive galaxies. The mass of the black hole is related to the properties of the host galaxy \citep{Magorrian1998,2013ARA&A..51..511K}, implying a close relation between the evolution of galaxies and the black hole at their centers. Mass accretion onto a SMBH is the mechanism that powers Active Galactic Nuclei (AGN) \citep{Salpeter1964,2008ARA&A..46..475H}, which are very powerful sources of X-ray radiation. X-ray emission in AGN mainly originates from a hot corona of relativistic electrons, located in the vicinity of the black hole. Thermal UV/optical photons emitted from the accretion disc are inverse-Compton scattered by the hot electrons into the X-rays, creating a power-law continuum (e.g., \citealt{1980A&A....86..121S}; \citealt{1993ApJ...413..507H}). Optically-thick geometrically-thin accretion discs can explain several key features of the observed Spectral Energy Distributions (SEDs) of AGN with moderate Eddington ratios ( $\lambda_{\rm Edd}=L_{\rm bol}/L_{\rm Edd}\in[0.01;0.3]$ , \citealt{Koratkar1999, Capellupo2015}). At higher accretion rates, the disc is expected to become geometrically thick or slim, and the nature of the accretion flow is expected to change dramatically by photon trapping through electron scattering in the dense matter and advection cooling (see \citealt{1988ApJ...332..646A} and \citealt{Wang2014a} for a recent review). Slim accretion discs are thought to have different properties from those of thin accretion discs, with the emitted radiation being significantly anisotropic and with a large fraction of the overall energetic output being carried away in the form of outflows. However, despite the recent advances in theoretical (e.g., \citealt{Jiang2019,Okuda_2021}) and observational (e.g., \citealt{Wang2014a,Du2018} and references therein) studies, Super-Eddington accretion is still the least understood accretion mode.
\begin{table*}
\caption{Summary of the 2020 simultaneous \textit{XMM-Newton} and\textit{ NuSTAR} observation of IRAS\,04416+1215. The Net count Rate is extrapolated between 0.3--10\,keV for EPIC cameras, 0.4--2\,keV for RGS and 3--25\,keV for both FPMA and B.}
\label{tab:observations_table}
\begin{tabular}{|c|ccc|cc|}
\hline
\hline
Telescope & \multicolumn{3}{c}{\textit{XMM-Newton}} & \multicolumn{2}{c}{\textit{NuSTAR}}\\
OBS.ID & \multicolumn{3}{c}{0852060101} & \multicolumn{2}{c}{60560026002} \\
Instrument & Epic-pn & Epic-MOS 1+2 & RGS 1+2 & FPMA& FPMB\\
Start Date &2020-02-17 & 2020-02-17 & 2020-02-17 & 2020-02-17 & 2020-02-17\\
Time (UT) & 18:24:20 & 18:16:01 & 18:15:18 & 16:16:09 & 16:16:09\\
End Date & 2020-02-18 & 2020-02-18 & 2020-02-18 &2020-02-18 &2020-02-18  \\
Time (UT) & 16:55:01 & 16:56:24 & 16:58:50 & 13:37:49 & 13:37:49\\
Exposure Time (ks) & 81.1 & 81.6 & 81.8 & 76.9 & 76.9 \\
Net Exposure Time (ks) & 56.5 & 78.5 & 16.3 & 71.3 & 71.3 \\
Net count Rate ($\rm{counts}/s$)& 0.477 & 0.116 & 0.019& 0.026 & 0.025\\
\hline
\hline
\end{tabular}
\end{table*}
Few models are available in literature of the emerging X-ray spectrum from high accretion rate sources. The common properties are the flattening spectrum at soft X-ray energies and the high energy cut-off almost independent from the accretion rate \citep{Wang2003}. Supercritical accretion flows produce radiation-pressure driven outflows, which will Compton up-scatter soft photons from the underlying accretion flow, making hard emission. Radiation-hydrodynamic simulations show that the Compton-$y$ parameter of the radiation pressure driven outflow from a super-Eddington accretion flow is around unity, thus the expected photon index of the primary power-law emission is $\sim 2$. \citep{King2003,2009PASJ...61..769K}.\\
A large fraction of AGN, classified in the optical as narrow-line Seyfert\,1 galaxies (NLS1s; \citealp{1994ApJ...435L.125M,2002A&A...388..771C,Collin_2004}), are widely believed to host SMBHs accreting close or above the Eddington limit \citep{Pounds1995,Komossa_2006,2016MNRAS.455..691J,10.1093/mnras/stx718}. The most common NLS1s defining criterion is the width of the broad component of their optical Balmer emission lines, in combination with the relative weakness of the [OIII]$\lambda$5007 emission \citep{1985ApJ...297..166O,2014Natur.513..210S}. NLS1 galaxies also show strong FeII emission, which, combined with their other properties suggests that these AGN contain relatively small supermassive black holes accreting at a rate close or above the Eddington limit \citep{1992ApJS...80..109B}. In the X-ray band, NLS1s show a rapid and large variability, with the break time scale having the behaviour observed in most AGN, increasing proportionally with black hole mass and decreasing with increasing accretion rate \citep{Uttley2005,McHardy2004,McHardy2006}. NLS1s show complex spectral properties as steep X-ray spectral slope ($\Gamma \sim 2.0 - 2.2$, e.g. \citealt {1997MNRAS.285L..25B,10.1046/j.1365-8711.1999.02811.x}) and evidence for cold and ionized absorption, partial covering, and strong features of reprocessed radiation, as well as a soft excess below 1 keV and a dip at $\sim$7 keV \citep{1996A&A...309...81W, 2000A&A...354..411K,Fabian_2002,2006MNRAS.365.1067C}. \citet{Pounds1995}, studying the spectrum of the NLS1 RE\,J1034+39, suggested that the steep spectrum of these objects could be the result of the much stronger Compton cooling of the corona by the strong radiation field from the super-Eddington disc, which may cause also a lower temperature of the AGN corona, as seen also by \citet{Kara_2017} in the NLS1 Ark\,564. Moreover, strong gas outflows are naturally expected during Super-Eddington accretion episodes \citep{Ballantyne2011,Zubovas2012,2014ApJ...796..106J} due to the intense radiation pressure associated with these events. The presence of outflowing disc winds has also been observed in some high-redshift QSOs accreting close to the Eddington limit \citep{Chartas2003, Lanzuisi2012, Vignali2015, Lanzuisi2016} and in ultraluminous X-ray sources (ULXs) \citep{2017AN....338..234P}. \\
Highly-accreting SMBHs are well known to show a prominent soft excess in the X-ray band. The origin of this feature is still debated. Two possible models have been debated in the past decade: a warm corona and blurred ionized reflection. In the former model  a warm ($kT \sim 0.5 - 1$ keV), optically-thick Comptonising layer above the accretion disc is believed to be responsible for this emission (e.g. \citealt{1998MNRAS.301..179M,2012MNRAS.420.1848D,2018A&A...611A..59P,2020A&A...634A..85P}).  In this framework the sharp spectral drop present in the spectra of NLS1s around $\sim$7 keV is interpreted in terms of relativistically blurred ionised reflection from the accretion disc, with the feature being associated to the blued edge of the relativistically-blurred iron emission line \citep{Fabian_2004}. Alternatively, this sharp spectral feature could be interpreted as the result of absorption through distant partial covering clouds \citep{Boller_2002}, or as a P-Cygni profile from emission and absorption in a Compton-thick wind \citep{2007MNRAS.374L..15D}.\\
It is well known that the X-ray emission of AGN is produced in a hot corona via thermal Comptonization of optical/UV seeds photons, emitted by the accretion disc \citep{Haardt1991,1993ApJ...413..507H}. Most of the AGN coronae observed so far have temperatures that lay close to the edge of the region in the compactness-temperature diagram which is forbidden due to run-away pair production \citep{Fabian2015, Fabian2017}.
The recently observed anti-correlation between the coronal temperature and the coronal optical depth \citep{2018A&A...614A..37T} suggests that some differences are required in the geometry of the accretion flow or in the intrinsic disc emission
for different sources in a configuration of radiative balance. In order to understand the accretion properties of AGN, it is essential to have good constraints on the coronal parameters that characterize the X-ray emission (i.e. the primary power-law photon index, $\Gamma$ and the high energy cut-off, $E_{\rm cut}$) and their relations with other parameters of the systems like the black hole mass or the Eddington ratio ($\lambda_{\rm Edd}$).
The $E_{\rm cut}-\lambda_{\rm Edd}$ relation is still under debate, recent work from \citet{Ricci_2018} on the \textit{Swift}/BAT AGN catalogue show a clear anticorrelation between the cut-off energy and the Eddington ratio with the sources accreting at lower Eddington ratio showing an higher cutoff and vice-versa a positive correlation in the relation $\Gamma-\lambda_{\rm Edd}$ \citep{2017MNRAS.470..800T,HUANG2020}. To obtain good measures of the coronal parameter it is crucial to disentangle the primary X-ray emission from the other spectral features, like the reflection component from cold circumnuclear material. Thanks to its unprecedented hard-band sensitivity Nuclear Spectroscopic Telescope Array (\textit{NuSTAR}, \citealt{Harrison2013}) mission, alone or with simultaneous observations with other X-ray observatories operating below 10 keV, such as the X-ray Multi-Mirror Mission (\textit{XMM-Newton}, \citealt{Jansen2001}), can provide strong constraints on the temperature of the AGN corona. Detailed broad-band spectroscopy using \textit{NuSTAR} has been performed on a large number of nearby AGN (e.g.,\citealt{2014ApJ...794..111B, 2014ApJ...788...61B, 2014ApJ...787...83M, Fabian2015, Tortosa_2016, Kara_2017, 2018A&A...614A..37T,2020ApJ...905...41B}), allowing to infer the high-energy cutoff and to disentangle X-ray continuum and reprocessed radiation.\\
Here we report the X-ray spectral and timing analysis of the joint \textit{XMM-Newton} and \textit{NuSTAR} observations of a IRAS\,04416+1215 a nearby (z = 0.0889, \citealt{Boller1992}) hyper-Eddington AGN. The source is part of a \textit{XMM-Newton}/\textit{NuSTAR} campaign that aims to constrain the broad-band X-ray properties of eight super-Eddington AGN from the best sample of bona-fide super-Eddington sources available, i.e. Super-Eddington Accreting Massive Black Holes (SEAMBHs, \citealt{Du2014,Wang2014a,2015ApJ...806...22D}) which contains exclusively objects with black hole masses estimated from reverberation mapping. In this campaign we are carrying out to study the broad-band X-ray properties of super-Eddington AGN, all the sources have new \textit{NuSTAR} observations performed simultaneously with \textit{XMM-Newton} or \textit{Swift-XRT}. IRAS\,04416+1215 has bolometric luminosity  $\log(L_{\rm bol}/\rm erg\,s^{-1})=47.55$, according to \citealt{2016MNRAS.458.1839C} and $\log(L_{\rm bol}/\rm erg\,s^{-1})=45.52$ according to \citealt{liu2021observational}. The former estimate is computed using, for the spectral energy distribution (SED) fitting procedure, the \citet{2012MNRAS.426..656S} code, including the comparison of the observed SED with various combinations of disc SEDs covering the range of mass, accretion rate, spins and taking into account the correction for intrinsic reddening and host-galaxy contribution. In the latter estimate, the SED fitting is done using the more semplicistic templates from \citet{Krawczyk_2013}. The dimensionless accretion rate \citep{Du2014} and black hole mass of the source are  $\log(\dot{\mathscr{M}}$)=$2.63^{+0.16}_{-0.67}$ and $\log(M_{\rm BH}/M_{\odot}$)=$6.78^{+0.31}_{-0.06}$ with the reverberation mapping technique \citep{2015ApJ...806...22D}, respectively, where $\dot{\mathscr{M}}\equiv\dot{M}_{\bullet}c^2/L_{\rm Edd}$, $\dot{M}_{\bullet}$ is mass accretion rates, $c$ is speed of light and $L_{\rm Edd}$ is the Eddington luminosity. The dimensionless accretion rate is estimated by $\dot{\mathscr{M}}=20.1\,\ell_{44}^{3/2}M_7^{-2}$ from the Shakura-Sunyaev disc model \citep{2015ApJ...806...22D}, where $\ell_{44}$ is the 5100\AA\, luminosity in units of $10^{44}\,{\rm erg\,s^{-1}}$ and $M_7=M_{\bullet}/10^7M_{\odot}$. This approximation is valid for $\dot{\mathscr{M}}\lesssim 10^3$. To compute the Eddington ratio we assumed the bolometric luminosity value from \citet{2016MNRAS.458.1839C}, which is a better and more trustable estimate of the bolometric luminosity of the source, obtaining $\lambda_{\rm Edd}\sim 472$. This value is in perfect agreement with the dimensionless accretion rate from \citet{Du2014}. However even assuming the luminosity from \citealt{liu2021observational}, with which the value of the accretion rate would be $\lambda_{\rm Edd}\sim 4.40$, the source would remain a super-Eddington accreting AGN.
IRAS\,04416+1215 turned out to be the most peculiar of our sample, it is classified as NLS1 galaxy, showing narrow H$\beta$ line (FWHM=$1670 \, \rm km \, \rm s^{-1}$, \citealp{Moran1996}) and very broad [OIII] (FWHM=$1150 \, \rm km \, \rm s^{-1}$, \citealt{Veron2001}) lines, which is typically found in sources accreting at such high Eddington accretion rates \citep{2005ApJ...627..721G,2009ApJ...699..638H}. The source shows a photon index in the Roentgen Satellite (\textit{ROSAT}) (0.1--2.4\,keV) energy band, of $\Gamma=2.96 \pm 0.50$ \citep{Boller1992} and of $\Gamma=2.46^{+0.27}_{-0.26}$ for the rest-frame $>2$\,keV spectrum, according to \citet{2021ApJ...910..103L}.\\
The paper is organized as follows. In Section \S \ref{sect:obs_data_reduction}, we present the X-Ray \textit{XMM-Newton} and \textit{NuSTAR} simultaneous observations. In Section \S \ref{sect:timing} and Section \S \ref{sect:spectralanaysis} we describe the timing and spectral data analysis processes. In Section \S \ref{sect:discussion} we discuss the results of our analysis which are summarized in Section \S \ref{sect:conclusion}.\\
Standard cosmological parameters (H=70\,km\,s$^{-1} \rm Mpc^{-1}$, $\Omega_{\Lambda}$=0.73 and $\Omega_m$=0.27) are adopted throughout the paper.\\
\begin{figure*}
	\includegraphics[width=\textwidth]{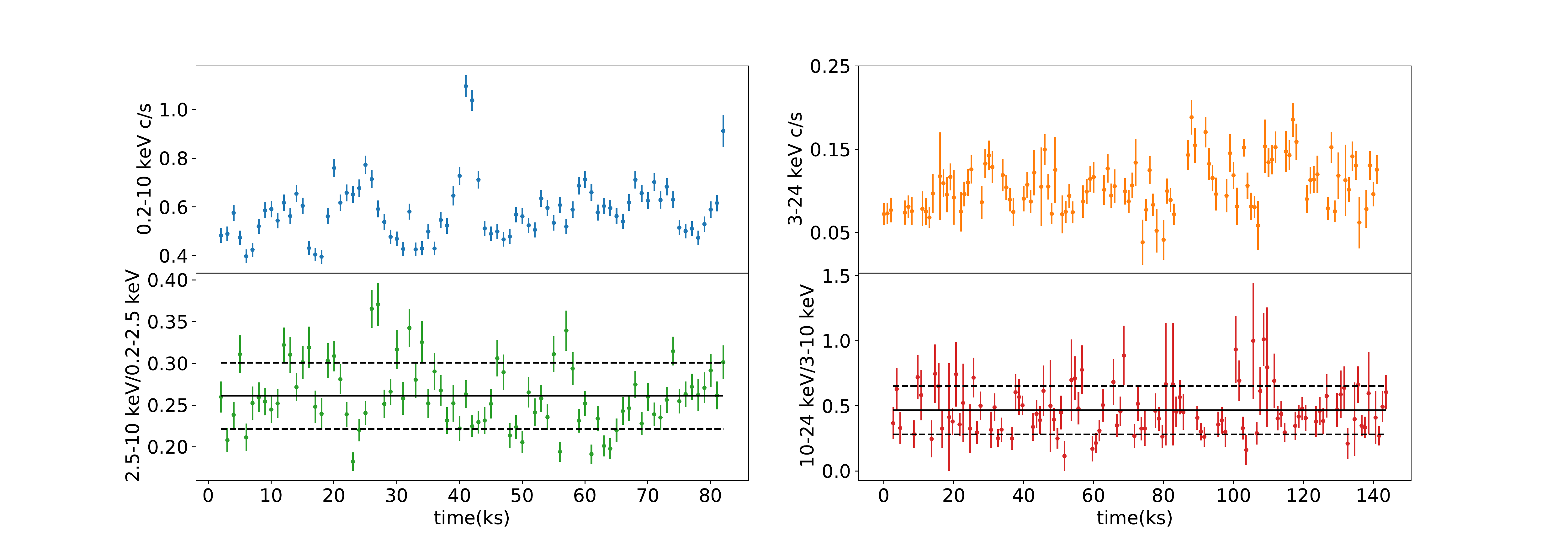}
    \caption{Top left panel: IRAS\,04416+1215 \textit{XMM-Newton} EPIC-pn background subtracted light curves (blue) in the: 0.2--10\,keV energy band (blue). Bottom left panel: ratio between the \textit{XMM-Newton EPIC-pn background subtracted light curves in the 2.5--10\,keV and 0.2--2.5\,keV energy bands (green)}. Top right panel: \textit{NuSTAR} FPMA+B background subtracted light curve in the 3--24\,keV energy band (orange). Bottom right panel: ratio between the \textit{NuSTAR} background subtracted light curves in the 10--24\,keV and 3--10\,keV energy bands (red).  All the light curves are extracted using a binning time of 1000\,s. Balck solid and dashed lines indicate the mean and the standard deviation respectively.}
    \label{fig:light_curves}
\end{figure*}
\section{Observations and data reduction}
\label{sect:obs_data_reduction}
IRAS\,04416+1215 was observed by \textit{NuSTAR} between 17-18 February 2020 simultaneously with \textit{XMM-Newton} (P.I. C. Ricci). The \textit{XMM-Newton} observation was performed with the European Photon Imaging Camera (EPIC hereafter) detectors, and with the Reflection Grating Spectrometer (RGS hereafter; \citealt{denHerder2001}). The EPIC cameras were operated in small window and thin filter mode. \textit{NuSTAR} telescope observed IRAS\,04416+1215 with its two coaligned X-ray telescopes Focal Plane Modules A and B (FPMA and FPMB, respectively). Details on duration and exposure of the observations are reported in Table \ref{tab:observations_table}.
\subsection{XMM-Newton}
The event lists of the EPIC cameras, both pn \citep{Struder2001} and MOS \citep{Turner2001}, are extracted with the \textsc{epproc} and \textsc{emproc} tools of the standard System Analysis Software (\textsc{SAS} v.18.0.0; \citealt{Gabriel2004}). The extraction radii and the optimal time cuts for flaring particle background were computed via an iterative process which maximizes the SNR, similar to the approach described in \citet{Piconcelli2004}. The spectra were extracted after checking that no significant pile-up affected the data, as indicated by the SAS task \textsc{epatplot}; the resulting optimal extraction radius was $30\arcsec$ and the background spectra were extracted from source-free circular regions with radii of $\sim$ $60\arcsec$ for both the EPIC and the two MOS. Response matrices and auxiliary response files were generated using the SAS tools \textsc{rmfgen} and \textsc{arfgen}, respectively. EPIC-pn spectra had a net exposure time of 56.5\,ks, the MOS spectra had both a net exposure time of 78.5\,ks. EPIC-pn and MOS spectra were binned in order to over-sample the instrumental resolution by at least a factor of three and to have no less than 20 counts in each background-subtracted spectral channel. EPIC-pn light curves are also extracted by using the same circular regions for the source and the background as the spectra.\\
RGS spectra are generated by using \textsc{rgsproc}, screening times with high particle background through the examination of the RGS light curves, and using \textsc{rgsfilter} and \textsc{rgsspectrum} to produce clean spectra.\\
It was not possible to perform a statistically meaningful analysis using RGS spectra because, even combining RGS1 and RGS2 spectra together with \textsc{rgscombine}, the spectrum had insufficient S/N. 
\subsection{NuSTAR}
The \textit{NuSTAR} Level 1 data products were processed with the \textit{NuSTAR} Data Analysis Software (\textsc{NuSTARDAS}) package (v.1.9.6) within the \textsc{heasoft} package (version 6.28). Cleaned event files (level 2 data products) were produced and calibrated using standard filtering criteria with the \textsc{nupipeline} task, and the latest calibration files available in the \textit{NuSTAR} calibration database (CALDB 20200813). For both FPMA and FPMB the radii of the circular region used to extract source and background spectra were $40\arcsec$ and $60\arcsec$, respectively; no other bright X-ray source is present within $40\arcsec$ from IRAS\,04416+1215, and no  source was present in the background region. 
The low-energy (0.2--5\,keV) effective area issue for FPMA \citep{Madsen2020} does not affect our observation, since no low-energy excess is found in the spectrum of this detector. The net exposure times after this process were 71.3\,ks for both FPMA and B. The two spectra were binned in order to over-sample the instrumental resolution by at least a factor of 2.5 and to have a Signal-to-Noise Ratio (SNR) greater than 3 in each spectral channel. Light curves are extracted using the \textsc{nuproducts} task, adopting the same circular regions as the spectra.

\section{Timing analysis}
\label{sect:timing}
First we investigated the temporal properties of the \textit{XMM-Newton} EPIC-pn observations of IRAS\,04416+1215. The EPIC-pn light curves showed the presence of rapid variability (see left panels of Figure \ref{fig:light_curves}). In particular there is a rapid increase followed by a decrease of the count rate by a factor two in less than two hours during the observation. This feature is present in the light curves extrapolated in different energy bands, but it does not seem to affect the ratio between the flux in the 2.5--10\,keV and the 0.2--2.5\,keV band, (see left panels of Figure \ref{fig:light_curves}).\\
To check whether flux variability affects the spectral analysis we extracted, together with the total time-averaged EPIC-pn spectrum, two spectra applying two different count-rate (CR) cuts: S1 and S2 (S1 with $c/s\geq 0.6$, and S2 with $c/s <0.6$). We looked for any spectral differences in these spectra by fitting all of them in the  0.3--10\,keV energy band using our best-fitting model (see Section \S \ref{sect:xmm}). We fitted all three spectra using the best-fitting model leaving the parameters free to vary for each spectrum, and we obtained a very good fit ($\chi^2$=1032 for 999 d.o.f). For both the spectra with the two different cuts in count-rate the values of the parameters are consistent within the errors with the  values obtained by fitting the total time averaged EPIC-pn spectrum. The only difference between the three spectra was in the 2--10\,keV flux: for the total time averaged EPIC-pn spectrum the flux was $1.13 \pm 0.20 \times 10^{-12}$ erg\,cm$^{-2}$\,s$^{-1}$ while for the S1 and the S2 spectra it was respectively $1.36 \pm 0.32\times 10^{-12}$ erg\,cm$^{-2}$\,s$^{-1}$ and $1.0 \pm 0.15\times 10^{-12}$ erg\,cm$^{-2}$\,s$^{-1}$. Since there were no substantial differences in the spectral shape and in the extrapolated parameters values in the three spectra, we decided to use the  total time averaged EPIC-pn spectrum in our analysis.\\
We investigated also the temporal properties of the \textit{NuSTAR} observation of IRAS\,04416+1215. Visual inspection of the \textit{NuSTAR} light-curves shows no evidence of strong flux and spectral variability. Since no spectral variation (less than 10\%) is found in the ratio between the 3--10 and 10--24\,keV count rates (see right panels of Figure \ref{fig:light_curves}), we decided to use time-averaged \textit{NuSTAR} spectra in the spectral analysis, to improve spectral statistics.
\subsection{Variability Spectrum}
\label{variability_spec}
\begin{figure}
	\includegraphics[width=\columnwidth]{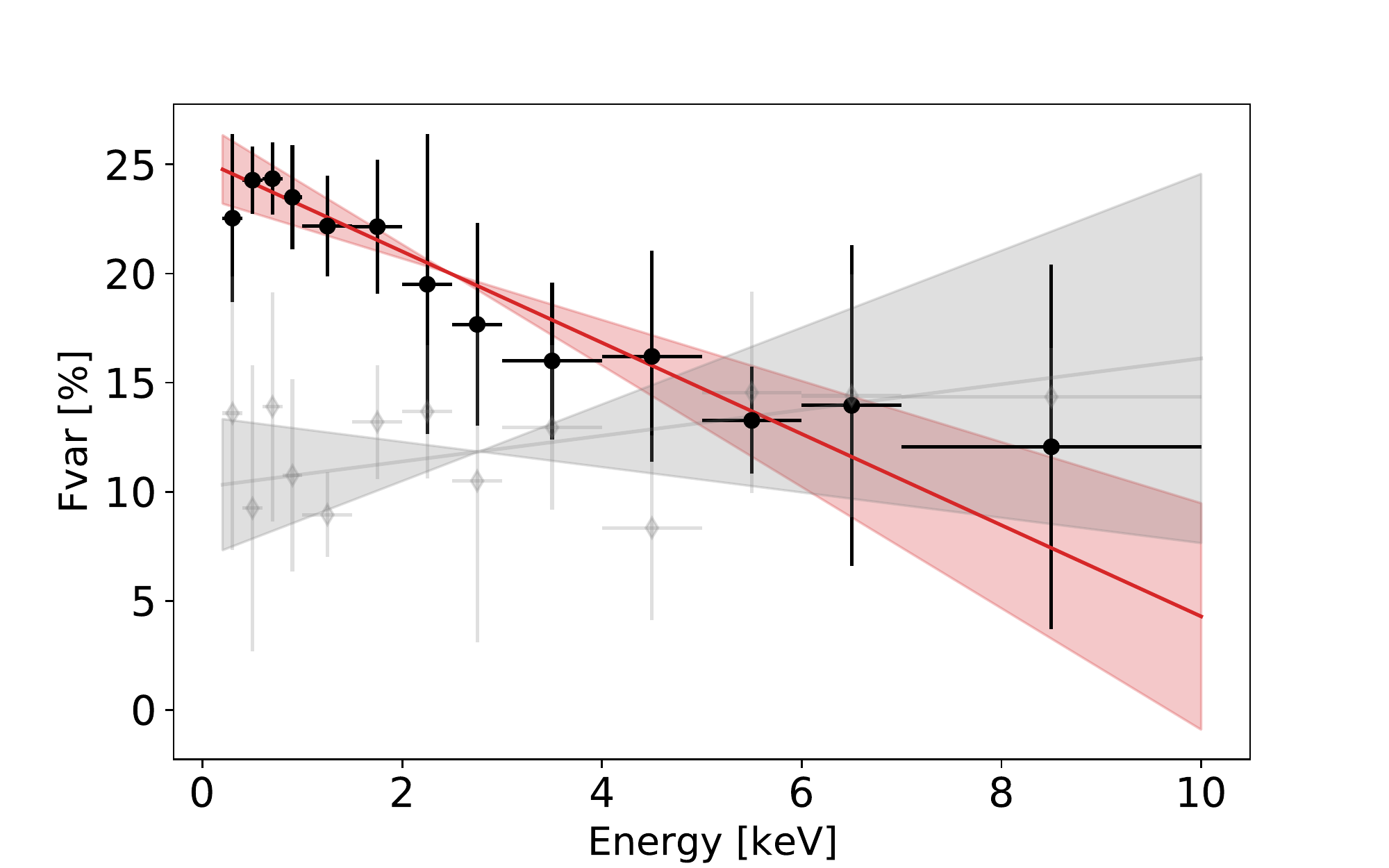}
    \caption{Fractional variability spectra for the EPIC-pn observation of IRAS\,04416+1215 (black) and median of the fractional variability spectra of the \textit{XMM-Newton} observations of the BASS sample of Seyfert\,1 with $\log(M_{\rm BH}/M_{\odot})\in [6.5-7.0]$ (gray). The red (and gray) solid lines and the shaded regions represent the linear fit and the 3-$\sigma$ error on the fitting relation, respectively.} 
    \label{fig:Fvar}
\end{figure}

We looked at the variability spectrum of IRAS\,04416+1215 using the normalized excess variance ($\sigma^2_{\rm NXS}$, \citealt{1997ApJ...476...70N,2002ApJ...568..610E}) and its square root: the fractional root mean square variability amplitude (F$_{\rm{var}}$, \citealt{2003MNRAS.345.1271V}) for the \textit{XMM-Newton} observation. The F$_{\rm{var}}$ is the difference between the total variance of the light curve and the mean squared error that is normalised for the average of the N flux measurements squared. Being N the number of good time intervals in a light curve, and $x_i$ and $\sigma_i$ respectively the flux and the error in each interval, F$_{\rm{var}}$ is defined \citep{2003MNRAS.345.1271V} as:
\begin{equation}
F_{\rm var}=\sqrt{\sigma^2_{\rm NXS}}=\sqrt{\frac{S^2-\overline{\sigma}^2}{\overline{x_i}^2}}
\label{eq:excess_var}
\end{equation}
Where:
$S^2=\frac{1}{N-1}\sum_{i=1}^{N}[(x_i-\overline{x_i})^2]$ is the sample variance, i.e. the integral of the PSD between two frequencies, and $ \overline{\sigma}^2=\frac{1}{N}\sum_{i=1}^{N}[\sigma_i^2]$ is the mean square error. We computed F$_{\rm var}$ using a set of 8 background-subtracted EPIC-pn light curve sections 10\,ks-long with an increasing energy binning, computing the median of the F$_{\rm var}$ in each energy bin of all these sections. We fitted the variability spectrum with a simple linear relation: $F_{\rm var}=a$[E/keV]$x+b$ obtaining the following slope and intercept values: 
\begin{equation} 
a=-2.09 \pm 0.23 ; b=25.19 \pm 0.37
\label{eq:params}
\end{equation}
The Pearson’s correlation coefficient is -0.78 while the probability of the data set appearing if the null-hypothesis is correct is $1.5\times10^{-3}$. Figure \ref{fig:Fvar} shows the EPIC-pn F$_{\rm{var}}$ as a function of the energy (variability spectrum) together with the median of the fractional variability spectra of the \textit{XMM-Newton} observations of the Swift/BAT AGN Spectroscopical Survey (BASS\footnote{\url{www.bass-survey.com}}) sample of Seyfert\,1 with $\log(M_{\rm BH}/M_{\odot})\in [6.5-7.0]$, for a total of 16 objects (Tortosa et al., in prep.). We extrapolated the fractional variability from the \textit{XMM-Newton} broad-band (0.2–10 keV) light curves of all the observations of the type 1 AGN present in the BASS sample using 10 ks-long light curve sections. IRAS\,04416+1216 shows a different trend respect to the median of the Seyfert\,1 sample from BASS with almost the same black hole mass as IRAS\,04416+1215, with a higher variability in the soft band, suggesting that the most variable components are the ones found at low energy, such as the soft excess and/or ionized absorption \citep{1997ApJ...476...70N,2012yCat..35420083P,10.1093/mnras/stu2618}. The reason for this difference most probably is related to the extreme accretion of IRAS\,04416+1215, while the Eddington ratio of the BASS sample of Seyfert\,1 with black hole mass $\sim 3\times10^6 -1\times10^7$ is in the range of $\lambda_{\rm Edd}\in [0.003-0.71]$ with a median value of 0.15.
\section{Spectral Analysis}
\label{sect:spectralanaysis}
\begin{figure}
	\includegraphics[width=\columnwidth]{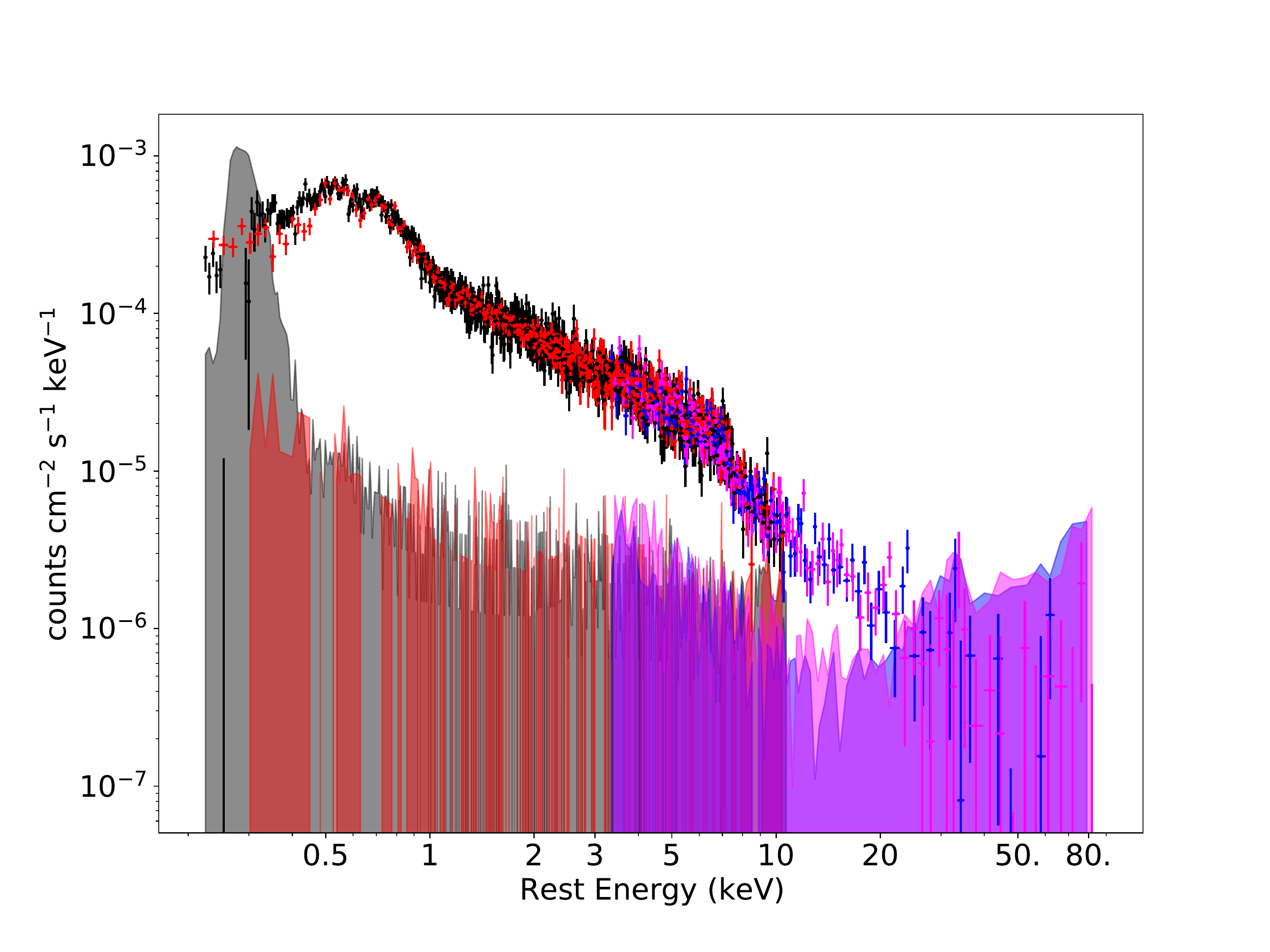}
    \caption{Background-subtracted EPIC-pn (black), MOS 1+2 (red), FPMA (blue) and FPMB (magenta) spectra in the 0.2--80\,keV range and the corresponding background levels of the simultaneous 2020 \textit{XMM-Newton} and \textit{NuSTAR} observation of IRAS\,04416+1215. The background contribution (shaded regions) dominates below $\sim 0.35$ keV and above $\sim$20\,keV. A spike in the EPIC-pn background above 8.5 keV is present but it does not contaminate the spectrum. Thus, we used the 0.35--24\,keV energy band for the X-ray analysis in this paper.}
    \label{fig:pn_mos_fpm}
\end{figure}
The spectral analysis has been performed with the \textsc{xspec} v.12.11.1b software package \citep{Arnaud1996}. Throughout the paper, all errors and upper/lower limits are calculated using $\Delta\chi^2$ = 2.71 criterion (corresponding to the 90\% confidence level for one interesting parameter), if not stated otherwise. The spectra obtained by the two\textit{ NuSTAR} modules (FPMA and FPMB) are fitted simultaneously, with a cross-normalization constant typically less than 5\% \citep{Madsen2015}. In the following fits, the Galactic column density at the position of the source ($N_{\rm H}=1.25 \times 10^{21} \rm cm^{-2}$, \citealp{HI4PI2016}) is always included, and modeled with \textsc{zTbabs} component \citep{Wilms2000} with $N_{\rm H}$ kept frozen to the quoted value. We also assumed Solar abundances, if not stated otherwise.\\
Fig. \ref{fig:pn_mos_fpm} shows the background subtracted EPIC pn, MOS 1+2 and FPMA-B spectra between 0.2-80 keV, plotted with the corresponding X-ray background. The spectra have been corrected for the effective area of each detector. The spectra are clearly background dominated at energies $E < 0.35$ keV and $E > 24$ keV so in this work, we will focus on the 0.35--24\,keV range for the spectral analysis.

\subsection{\textit{XMM-Newton} data analysis}
\label{sect:xmm}
\begin{figure}
	\includegraphics[width=\columnwidth]{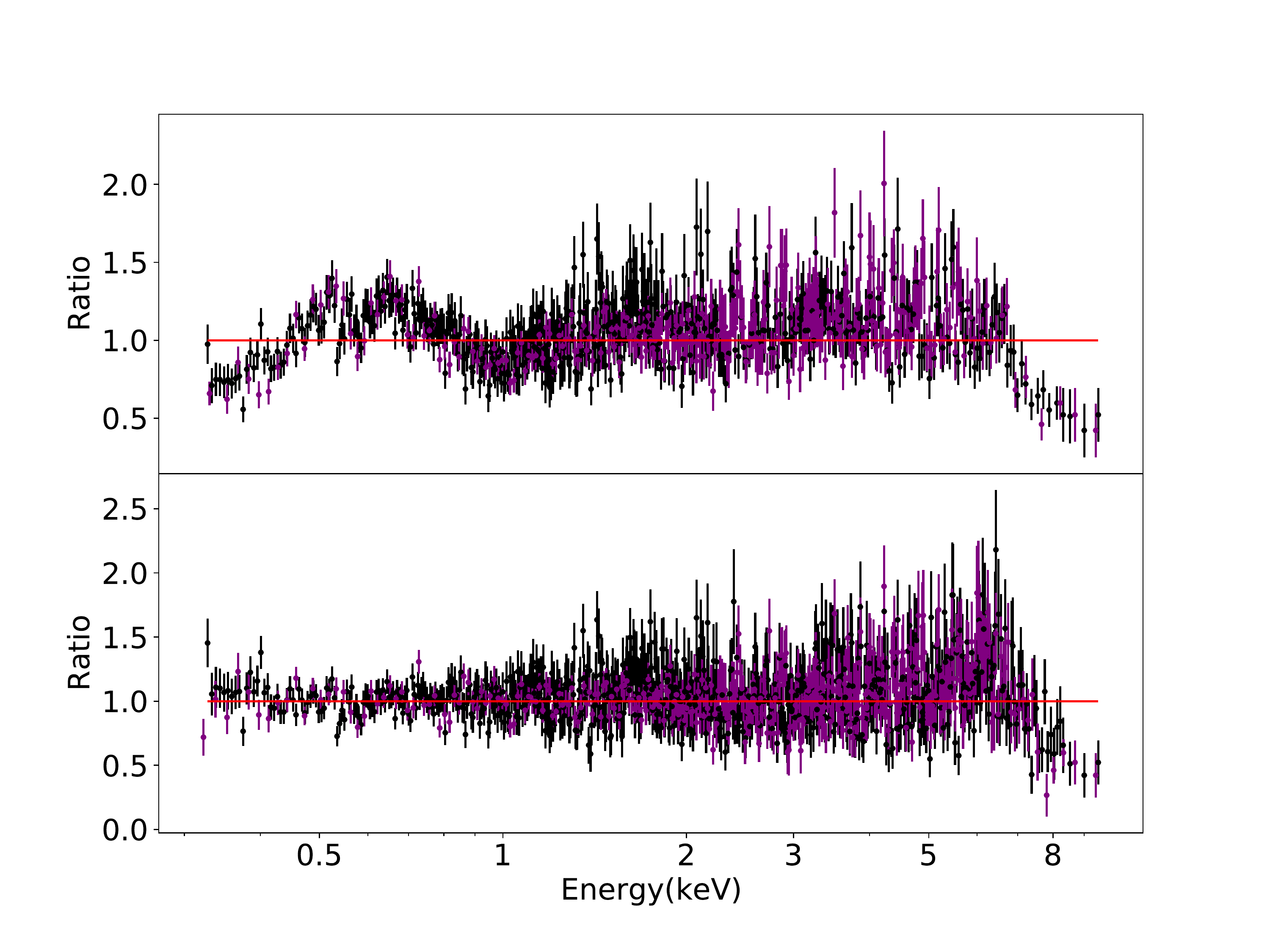}
    \caption{Ratio residuals for the EPIC-pn  (black) and MOS1+2 (purple). The model used consists of a power law for the primary continuum, a Gaussian line for the iron K$\alpha$ emission and a soft excess (top panel) and of a power law for the primary continuum, a Gaussian line for the iron K$\alpha$ emission, a soft excess component, a neutral absorption component and three ionized absorption components (bottom panel).}
    \label{fig:first_fit}
\end{figure}

We started our data analysis by fitting the 0.35--8.5\,keV \textit{XMM-Newton} EPIC-pn and MOS 1+2 spectra simultaneously with a phenomenological baseline model, composed of a power law for the primary continuum, a narrow Gaussian line at 6.4 keV, corresponding to the neutral iron K$\alpha$ emission line, which is a typical feature in Seyfert galaxies \citep{1994MNRAS.267..974N} and a soft excess modeled by a black-body component, absorbed by the Galactic column density. We tied all the MOS parameters to the pn values, apart from the normalization of the various components.  We found the photon index of the primary power law being too flat for a such rapidly accreting source ( $\Gamma=1.29 \pm 0.02$), as previous studies reported a positive correlation between photon-index and Eddington ratio \citep{Porquet2004,Piconcelli2005,Shemmer2006,Shemmer2008,Ricci_2017,2017MNRAS.470..800T}. This yielded a fit with $\chi^2$ = 1675 for 983 degrees of freedom (dof), which indicates that the model fails to properly reproduce the spectral shape of the source. Moreover clear residuals below 1\,keV suggested the presence of at least two absorption components affecting the soft X-rays (see Figure \ref{fig:first_fit}). 
To take into account these absorption components we included in the fitting model two \textsc{zxipcf} models for partial covering of partially ionized absorbing material. This model uses a grid of \textsc{xstar} \citep{Kallman2001} photionized absorption models (calculated assuming a turbulent velocity of 200\,$\rm km\,s^{-1}$) for the absorption, assuming that the absorber only covers some fraction of the source. We included also an additional neutral partial covering absorption component (\textsc{zpcfabs}). The fit slightly improved ($\chi^2/\rm dof=1179/975=1.21$). Since the covering fraction of the neutral absorption component was unconstrained, showing an upper limit of Cvf $>0.85$, we replaced it with a fully covering neutral absorption component (\textsc{zTbabs}). With this model we obtained a fairly good estimation of the photon index of the primary continuum of $2.03\pm0.08$, and the fit significantly improved ($\chi^2/\rm dof=1079/976=1.12$), although some residuals were still present below 1\,keV, suggesting that another absorption component was needed. We therefore added another \textsc{zxipcf} component. This procedure yielded a good fit ($\chi^2/\rm dof=1069/971=1.10$). All absorption components showed a covering fraction $\sim 1$. The observed redshift of the first absorber, which is comparable with the cosmological one of the source, is typical of Warm Absorbers (WAs) \citep{Blustin2005}. The second and the third absorbers showed redshifts ($z_{\rm obs,2}=-0.12 \pm 0.06$ and $z_{\rm obs,3}=-0.16 \pm 0.05$) more similar to Ultra Fast Outflows (UFOs, \citealt{Tombesi2013}).\\
\begin{figure}
	\includegraphics[width=\columnwidth]{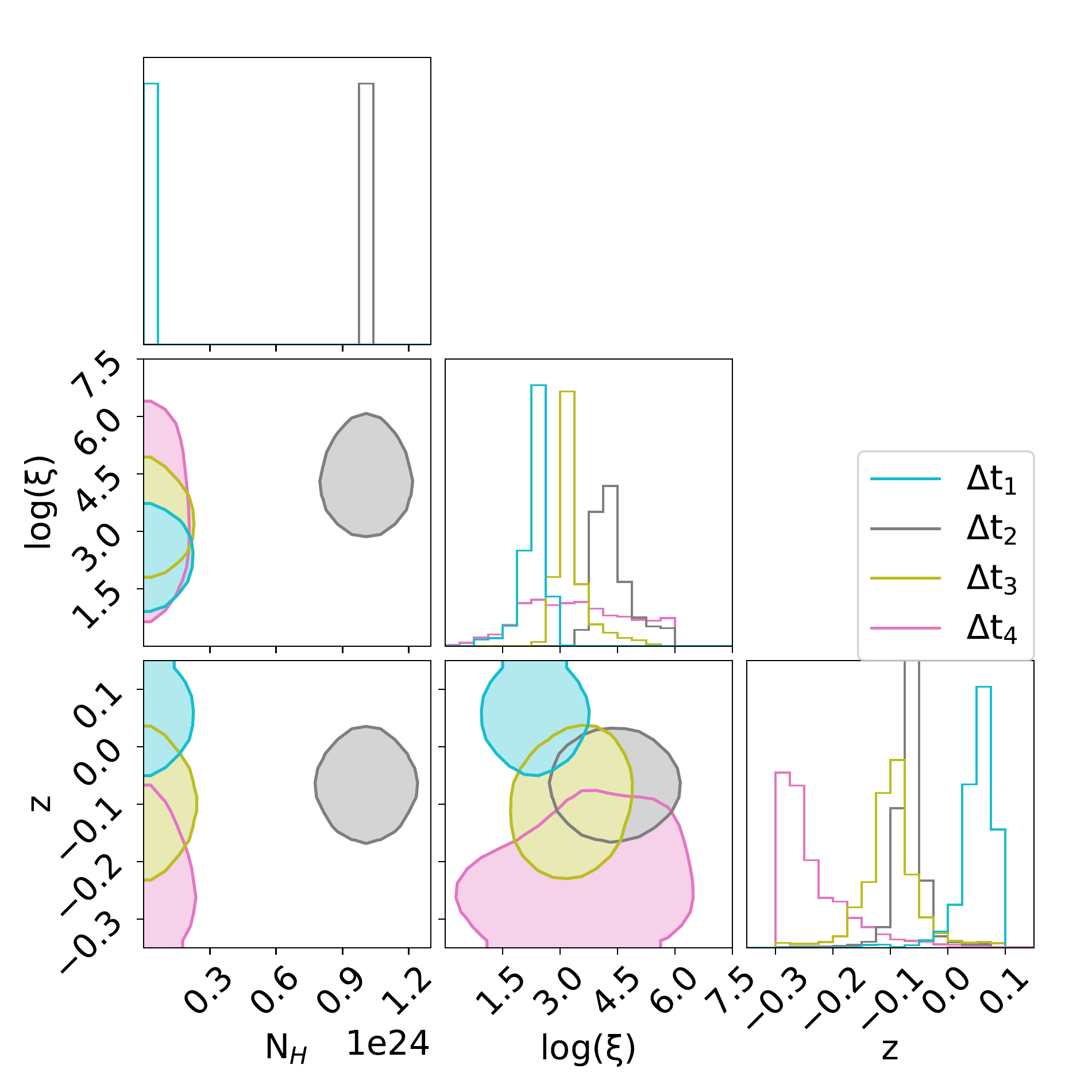}
    \caption{90\% contour plots resulting from MCMC analysis of the ionized absorption models applied to the \textit{XMM-Newton} spectra extrapolated over 20\,ks time intervals for Wind 3. We show the outputs for the ionization parameter (log($\xi$) [log(erg cm s$^{-1}$]), the column density ($N_{\rm H}$) and the observed redshift ($z$) for each time interval.}
    \label{fig:time_cuts_contours}
\end{figure}
For a more refined modeling of these absorbers we replaced the \textsc{zxipcf} components with detailed grids computed with the photoionization code \textsc{xstar}, with a spectral energy distribution described by a power law with a photon index of $\Gamma = 2$. These tables consider standard solar abundances from \citet{Asplund2009}, and take into account absorption lines and edges for all the metals characterized by an atomic number $Z \leq 30$. Following the previous consideration about the values of the parameters of the absorbers obtained with the \textsc{zxipcf} model, and considering typical values of turbulent velocity for WAs \citep{Laha2014}, we used one table computed considering a turbulent velocity of 100\,$\rm km\,s^{-1}$ for the first absorber (hereafter Wind 1) and two tables computed considering a turbulent velocity of 1000\,$\rm km\,s^{-1}$ for the second and the third absorbers (hereafter  Wind 2 and Wind 3) which is consistent with the typical value used for UFOs (e.g. \citealt{Tombesi2011,Gofford2013}). We included also a neutral absorption component (\textsc{zTbabs}), as we did before. The best-fitting model \footnote{\textsc{xspec} model: zTbabs * mtable(xout-mtable-v100.fits) * mtable(xout-mtable-v1000.fits) * mtable(xout-mtable-v1000.fits) * zTbabs * (powerlaw+zgauss+bbody)} was composed of a black body component for the soft excess, a Gaussian line for the iron k$\alpha$ emission line, a power law component for the primary continuum, a neutral absorption component and three ionized absorption components, one with turbulent velocity of 100\,$\rm km\,s^{-1}$ and two with turbulent velocity 1000\,$\rm km\,s^{-1}$. The iron K$\alpha$ emission line appears to be weak showing a flux of $8.47 \pm 0.55 \times 10^{-7}\,\rm ph\, \rm cm^{-2}\, \rm s^{-1}$ and an equivalent width of $46\pm13$ eV. The weak Fe K$\alpha$ line together with the presence of the sharp drop around $\sim$7\,keV, typical of NLS1 galaxies \citep{Boller_2002}, suggests that the X-ray spectrum may be reflection dominated \citep{Fabian_2002}. This will be investigated in details in the next section \S \ref{sect:bbanalysis}. In overall, this model provided a very good description of the data: $\chi^2$ = 998 for 967\,dof and a photon index of the primary continuum of $\Gamma = 1.94 \pm 0.17$ consistent with previous studies of super-Eddington sources \citep{1997MNRAS.285L..25B,Brightman2013}. However, some residuals above 6\,keV are still present (see bottom panel of Figure \ref{fig:first_fit}), suggesting the presence of a curvature related to a reflection component and/or to a cut-off at high energy. In the next section \S \ref{sect:bbanalysis} we exploit the \textit{NuSTAR} high sensitivity at high energy together with \textit{XMM-Newton} to describe this component with the analysis of the broad-band spectrum of IRAS\,04416+1215.\\
To search for possible variability of the absorption components, we also divided the EPIC-pn  \textit{XMM-Newton} observation in four time intervals of 20\,ks each. We analyzed the spectra for each interval to see if, during the observing time, there were any temporal variations in the absorbers. We applied the best-fitting model to all the spectra, keeping all the parameters free to vary during the fitting process. The resulting parameters and errors for the three wind components are shown in Table \ref{tab:timecuts}. We found that only Wind 3 can be considered variable within the errors and we show in Figure \ref{fig:time_cuts_contours} the 90\% Wind 3 contour plots of the ionization parameter, the column density and the observed redshift for each time interval. The possible variability of this WA component suggests that probably it can be intermittent and/or clumpy \citep{Tombesi2013,2018ApJ...864L..27F}.\\
Analysing the \textit{XMM-Newton} observation, we found that IRAS\,04416+1215 shows a 2--10\,keV luminosity of $L_{2-10}=2.60 \pm 0.78 \times 10^{43}$ $\rm erg\,s^{-1}$.
\begin{figure*}
    \includegraphics[width=\columnwidth]{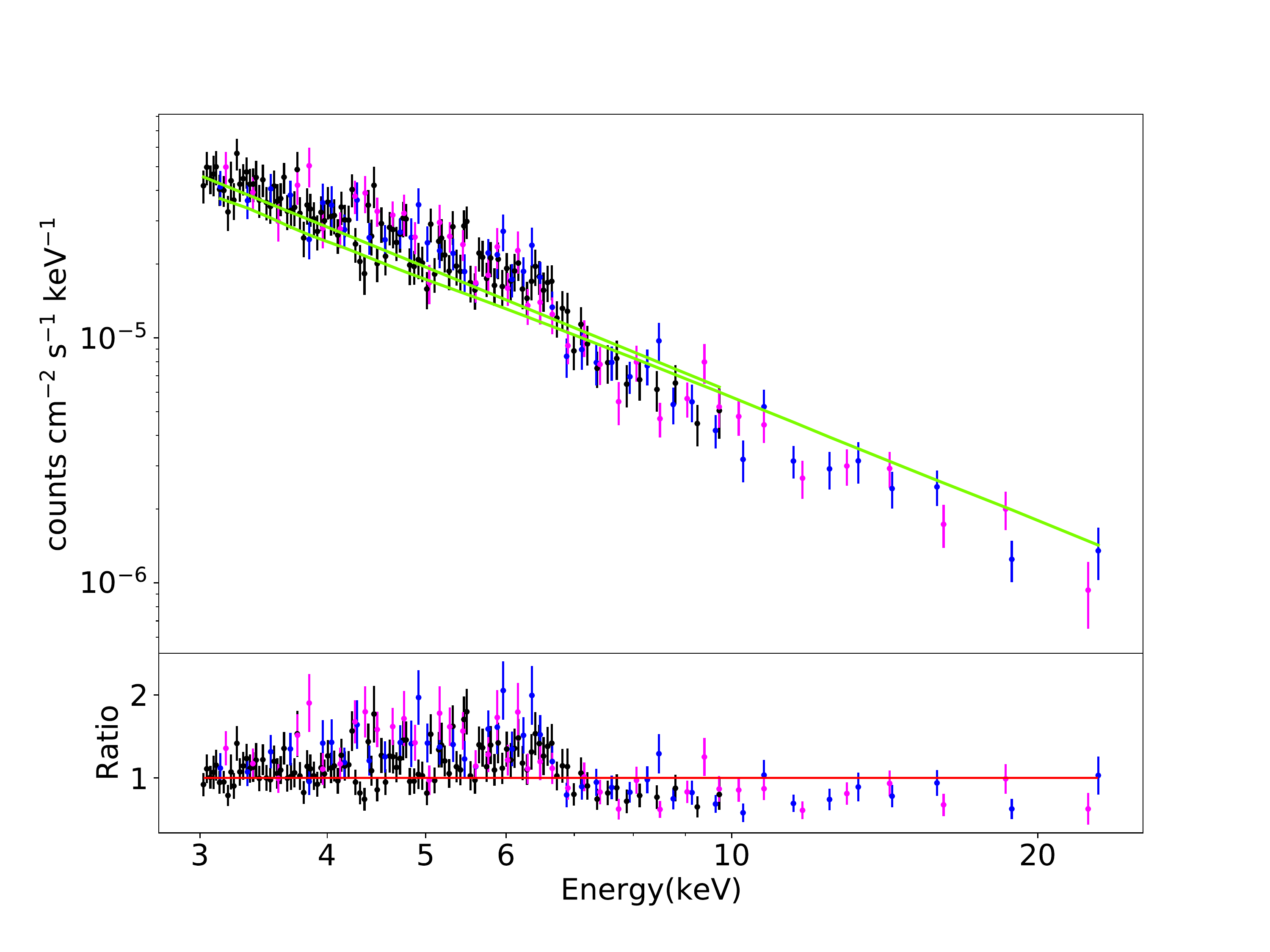}
    \includegraphics[width=\columnwidth]{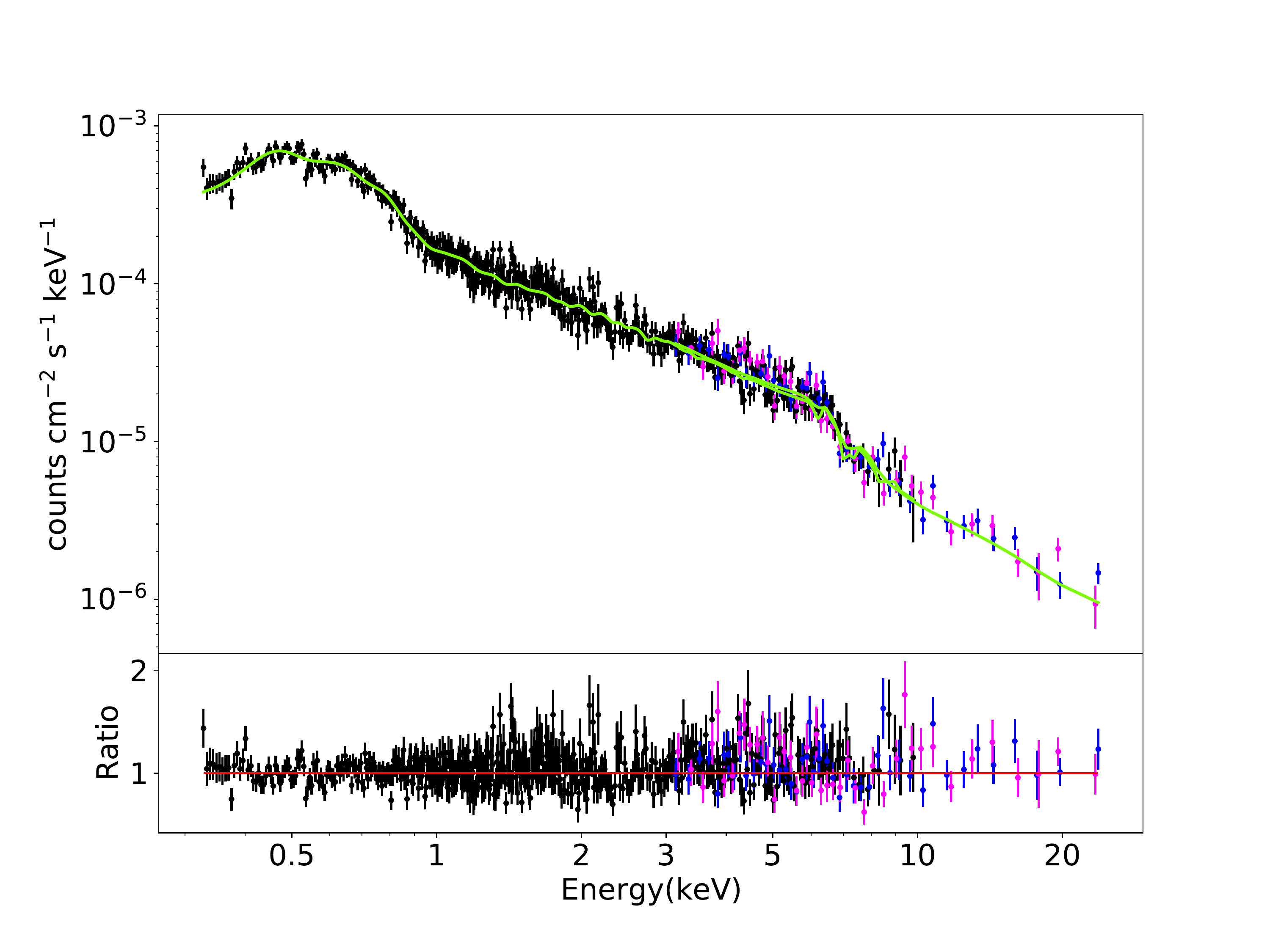}
    \caption{Data, fitting model (top panel) and residuals ratio (bottom panel) for EPIC-pn (black) and \textit{NuSTAR} FPMA (blue) and FPMB (magenta) spectra of the hyper-Eddington source IRAS\,04416+1215 in the range of 3--25\,keV considering the a baseline model composed by a power-law ($\Gamma=1.8$) with Galactic absorption (left panels) and considering the best-fitting model including the \textsc{xillver} reflection as reported in the Section \S \ref{sect:bbanalysis} (right panels).}
    \label{fig:baseline}
\end{figure*}
\subsection{Broad-Band data analysis}
\label{sect:bbanalysis}
Before analyzing the broad-band spectra, we first focused on the primary emission and on its reflected component using \textit{XMM-Newton} EPIC-pn and \textit{NuSTAR} FPMA and FPMB data, ignoring tha data below 3\,keV. During the fitting process we left all parameters, except the fluxes of the various components, tied together. The \textit{XMM-Newton} and the \textit{NuSTAR} FPMA calibration constants are fixed to 1.0 since any mismatch between the two instruments cannot be separated from intrinsic variation, while we left the \textit{NuSTAR} FPMB cross-calibration constant free to vary. First we fitted the data with a simple baseline model composed of a power-law absorbed by the Galactic absorption. This very simple test yielded a poor fit ($\chi^2=577$ for 402 dof). As shown in left panel of Figure \ref{fig:baseline}, the need for a more complex model to fit the data, as well as the presence of a cutoff at high energies, is evident. Thus, we tested the presence of the high energy cutoff together with a reflection component. Being this a first test, we modeled the primary continuum and the reprocessed emission testing only the standard reflection using the photoionized reflection model \textsc{xillver} version [1.4.3] \citep{Garcia2013}, which accounts also for the Fe K$\alpha$ emission line. The inclination angle was fixed to a value of $30^{\circ}$. The fit was good ($\chi^2=430$ for 379 dof). We found a rather high value of the iron abundance ($A_{\rm Fe}=1.81^{+1.15}_{-0.73}$) and a lower limit on the reflection fraction ($R_{\rm refl}>3.0)$, which suggests that the X-ray spectrum of IRAS04416+1215 could be reflection-dominated. The primary continuum showed a power-law ($\Gamma=1.76\pm0.09$) and a cut-off at $\sim50$\,keV.\\
Then, we extended the analysis including the whole 0.35--10\,keV \textit{XMM-Newton} EPIC-pn spectrum and the 3--24\,keV \textit{NuSTAR} FPMA and FPMB spectra. Following the previous steps of the analysis we adopted a fitting model composed of a black body, three ionized absorption components, one with turbulent velocity of 100\,$\rm km\,s^{-1}$ (Wind1) and two with turbulent velocity 1000\,$\rm km\,s^{-1}$ (Wind2 and Wind3). In all the fits we performed the value found for the \textit{NuSTAR} cross-correlation constant between FPMA and FPMB is 0.99$\pm$0.06. This value is consistent with what expected (e.g., \citealt{2015ApJS..220....8M}).\\
Given the complex spectral shape of the source we tested all the different flavours of \textsc{xillver} and \textsc{relxill} version [1.4.3] models \citep{Garcia2014, Dauser2014}. We tested also the AGN Super-Eddington accretion model \textsc{agnslim} \citep{2019MNRAS.489..524K}. For the sake of brevity, we report in this section only the best fit results but the detailed fitting procedure with all the different models is reported in Appendix \ref{app:fitting}.\\
\begin{figure*}
\centering
	\includegraphics[width=0.68\columnwidth]{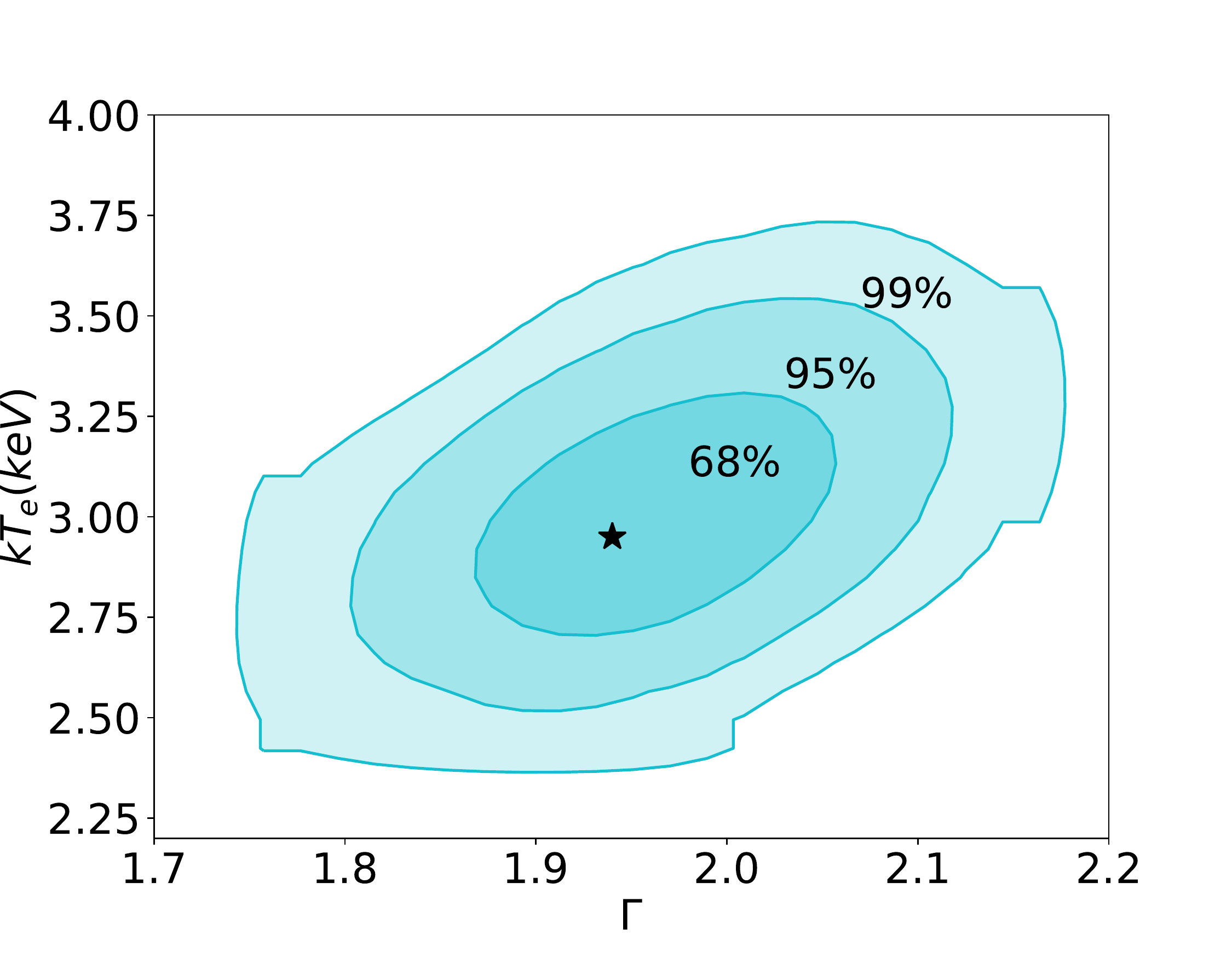}
	\includegraphics[width=0.68\columnwidth]{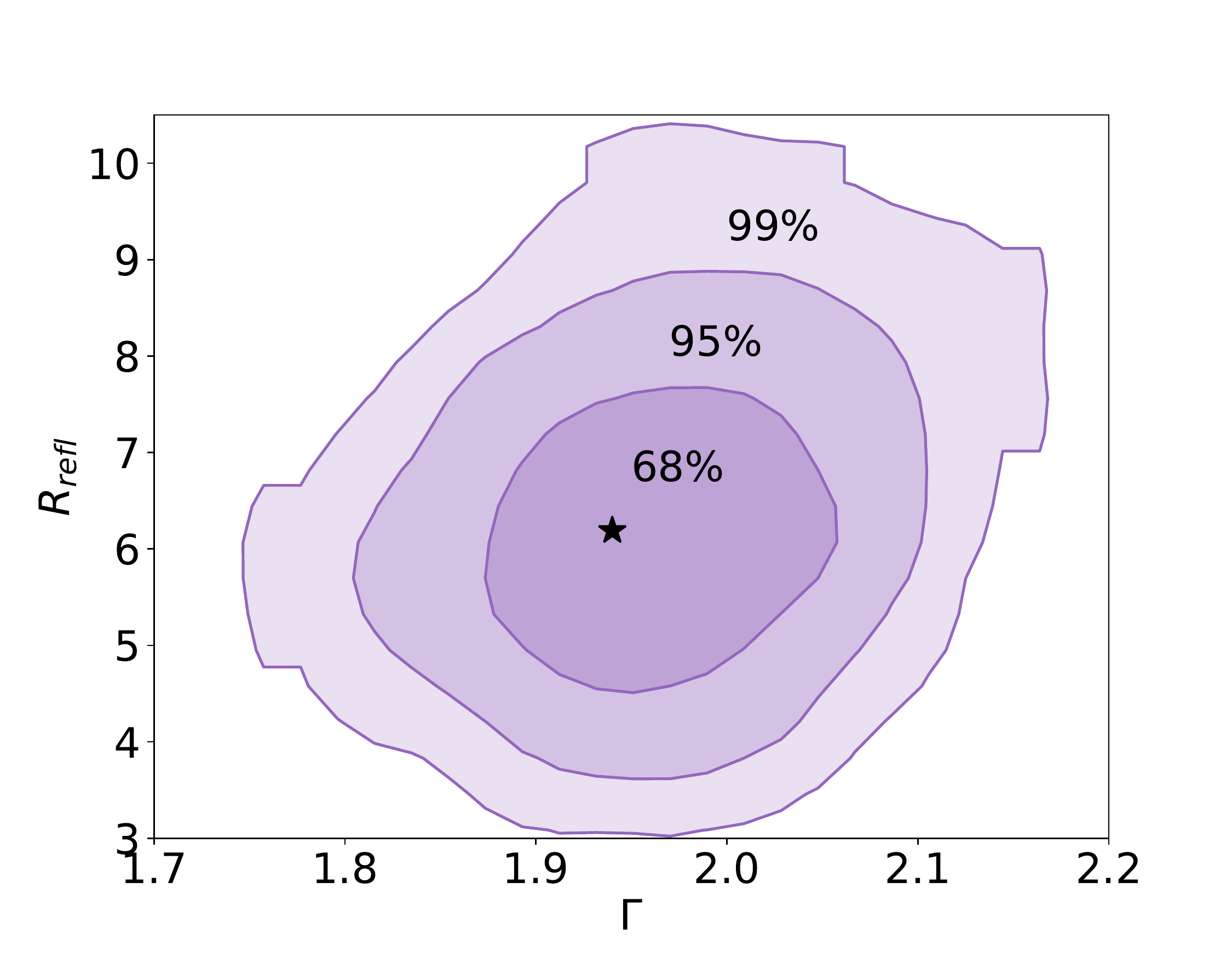}
	\includegraphics[width=0.68\columnwidth]{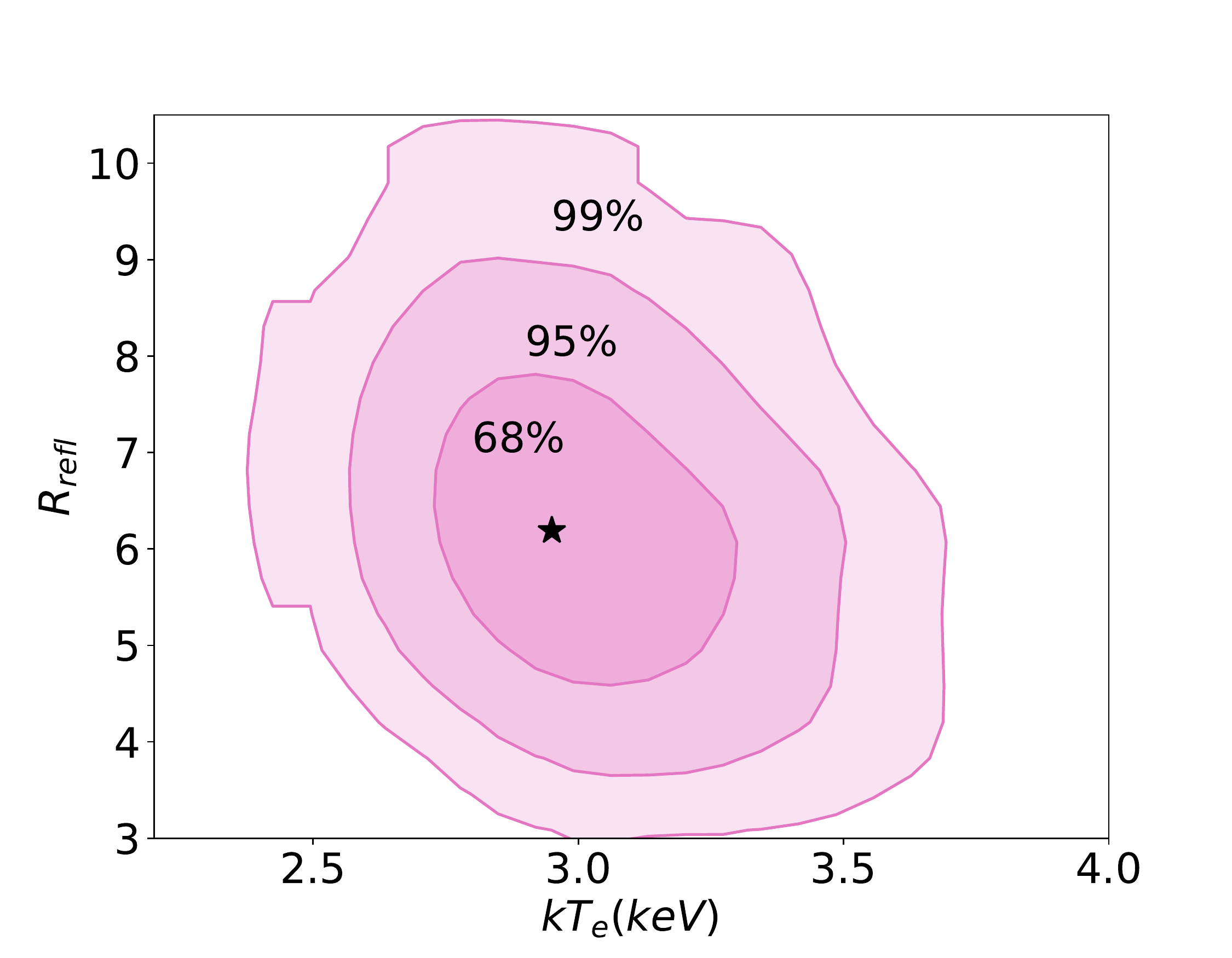}
    \caption{$kT_e$ vs $\Gamma$ contour plot (left panel), $R_{\rm refl}$ vs $\Gamma$ contour plot (midle panel) and $R_{\rm refl}$ vs $kT_e$ (right panel) resulting from the MCMC analysis of the \textsc{xillverCp} model applied to the broad-band \textit{XMM-Newton} and \textit{NuSTAR} 0.35--24\,keV spectra of IRAS\,04416+1215. The black star represents the best fit value of the parameters.}
    \label{fig:contours}
\end{figure*}
The reflection models including the relativistic reflection and/or allowing for a highest density of the accretion disc appeared to be statistically as good as the fit with the standard \textsc{xillver} model, but they showed some weaknesses. In the fit with the \textsc{relxill} model we did not find a good measurement for the spin of the black hole, finding only a lower limit in both in the case of standard relativistic reflection and of relativistic reflection from the high density accretion disc. Moreover, in the models for standard and relativistic reflection allowing a higher density for the accretion disc, the high-energy cut-off is fixed at 300\,keV but in all the previous fits we found a much lower values for the high energy cut-off of IRAS\,04416+1216. For these reasons we found these solutions unlikely and we believe the model with the standard \textsc{xillver} table being the best-fitting model\footnote{\textsc{xspec} model: constant * zTbabs * mtable(xout-mtable-v100.fits) * mtable(xout-mtable-v1000.fits) * mtable(xout-mtable-v1000.fits) * zTbabs * (xillver+bbody)} for IRAS\,04416+1215 (see right panels of Figure \ref{fig:baseline}).\\
In all the models tested we found an iron overabundance and a high value of the reflection fraction (see Table \ref{tab:bb_fit}), confirming the result obtained by fitting only the 3--24\,keV energy range, which indicates that IRAS04416+1215 has a reflection-dominated spectrum. If the accretion disc is not perfectly flat, but it is clumpy or has the shape of deep rings or sheets, the reflection component can be be stronger and the reflection fraction can be $>1$ \citep{Fabian_2002}. In particular this could happen when the disc is radiation-pressure dominated (e.g., \citealt{1974ApJ...187L...1L,1998ApJ...498L..13K,2001xeab.confE..71T}). Such a situation is most likely to occur when the accretion rate is close or above the Eddington limit, as in the case of the NLS1 galaxy 1H\,0707-495 \citep{2009Natur.459..540F} and of the Seyfert 2 galaxy IRAS\,00521-7054 \citep{2014ApJ...795..147R}. Similarly to 1H\,0707-495, IRAS\,04416+1215 shows a deep drop in the spectrum at almost the exact energy of the neutral iron edge, typical of many other NLS1 galaxies \citep{Boller_2002, 8202723}. This feature could be due to the blue wing of a line partially shaped by relativistic Doppler shifts, or to a photoelectric absorption edge due to the source being partially obscured by a large column of iron-rich material. The former requires a reflection fraction factor between 5 and 10, the latter requires an iron overabundance about 30 times the Solar value to fit this sharp feature in the spectrum at around $7$\,keV. In the analysis of IRAS\,04416+1215 we found a reflection fraction $>8$ and an iron abundance of $\sim 5$ which is well below the required overabundance to justify the partial covering model, supporting the hypothesis of the presence of a very strong X-ray reflection in the innermost regions of the source. 
\subsection{Comptonization features}
\label{sect:comptonization}
The coronal plasma electron temperature ($kT_e$) is expected to be related to the cut-off energy by $E_{\mathrm{cut}}=2-3\,kT_e$ \citep{Petrucci2000, Petrucci2001}. Following this relation applied to the value of the cut-off energy from the best fitting model (i.e, 44\,keV, see Table \ref{tab:bb_fit}) we derived the coronal temperature, which is $kT_e \sim 15 - 22$\,keV. Moreover, to directly measure the coronal temperature parameter, assuming that the primary emission is due to the Comptonization of thermal disc photons in a hot corona, we used the \textsc{xillverCp} and \textsc{relxillCp} versions of the tables in which the reflection spectrum is calculated by using a more physically motivated primary continuum, implemented with the analytical Comptonization model \textsc{nthcomp} \citep{Zdziarski1996, Zycki1999}, instead of a simple cut-off power law. In this model the seed photon temperature is fixed at 50\,eV. We found very good fits from the statistical point of view, with $\chi^2$=853 for 808 dof using the standard reflection table \textsc{xillverCp} and $\chi^2$=855 for 808 dof using the relativistic reflection table \textsc{relxillCp}. In both cases we obtained a photon index of $\Gamma \sim 1.98$ and a coronal temperature of $kT_e \sim 2.67$\,keV. These fits seem to be unlikely because the coronal temperature appears to be extremely low. Then we fixed the iron abundance, which was unconstrained and found to be A$_{\rm Fe}<0.65$, to the solar value, obtaining similar fit in terms of statistic ($\chi^2$=854 and $\chi^2$=864 for 808 dof, respectively for the \textsc{relxillCp} and the \textsc{xillverCp} table). We found a photon index $\Gamma = 1.88 \pm 0.04$, consistent with that of super-Eddington galaxies, while the coronal temperature was still extremely low (all the fitting parameters are reported in Table \ref{tab:bb_fit}). In the left panel of Figure \ref{fig:contours} the contour plot of coronal temperature ($kT_e$) versus the photon index of the power law ($\Gamma$) is shown, while in the middle panel we show the contour plot of the reflection fraction ($R_{\rm refl}$) versus the photon index of the power law and in the right panel is shown the contour plot of the reflection fraction versus the coronal temperature. This confirms that the value of the electron temperature, and that of the reflection parameters, are rather extreme. We furthermore tested a beta-version of the \textsc{xillver} and \textsc{relxill} tables allowing a higher density for the accretion disc and with the coronal temperature as free parameter: \textsc{xillverDCp} and \textsc{relxillDCp}\footnote{Courtesy of Javier García.}. Fitting with this new models we obtained good fits ($\chi^2_r=\chi^2/ \rm dof=1.08$ when using \textsc{xillverDCp} table and $\chi^2_r=\chi^2/ \rm dof=1.04$ when using \textsc{relxillDCp} table) but the relevant parameters are pegged to upper or lower limits (see Table \ref{tab:bb_fit}). 
\section{Discussion}
\label{sect:discussion}
\subsection{Location and energetic of the X-ray winds}
\label{sect:winds}
During the fitting process we took into account the absorption including three \textsc{xstar} tables as described in Section \S \ref{sect:xmm}. The resulting parameters for the absorption components are listed in Table \ref{tab:bb_fit}. The fitting parameters of these three wind components are constrained by the current data-set (except for the ionization parameter of the Wind 1, for which we obtained only an upper limit). Since the values of all the wind parameters are almost consistent with each other within the errors when considering different baseline models, we used the parameters obtained from the fit with  \textsc{xillver} model (see Table \ref{tab:bb_fit}) to compute the physical quantities related to the absorbers. First we derived the outflow velocity of the winds components. 
\begin{table}
\centering
\caption{Column density ($N_{\rm H}$), ionization fraction ($\log(\xi)$), observed redshift ($z_{\rm obs}$), Doppler shift ($z_{\rm w}$), velocity ($v_{\rm w}$, where c is the speed of light) and radial location for each absorber. We also reported the reduced chi square and the F-statistic null-hypothesis probability when adding each absorbing component to the fitting model described above.}
\label{tab:doppler_shift}
\begin{tabular}{lccc}
\hline
\hline
Parameter & Wind 1 & Wind 2 & Wind3\\
\hline
\hline
$N_{H} (10^{22} \rm cm^{-2})$ & $0.8\pm0.1$ & $2.8_{-3.1}^{+1.2}$ & $2.2_{-0.7}^{+1.1}$\\
$\log(\xi/\rm erg\,\rm s^{-1} \rm cm)$ &$0.13_{-0.05}^{+0.03}$ &$2.68_{-0.07}^{+0.08}$&$2.41_{-0.31}^{+0.13}$\\
$z_{\rm obs}$&$0.015_{-0.021}^{+0.016}$ & $-0.05_{-0.07}^{+0.01}$&$0.06\pm0.01$\\
$v_{\rm w}/c$&$0.07\pm0.01$&$0.13\pm0.03$&$0.025\pm 0.008$\\
$r_{\mathrm{min}} [10^{-4} \mathrm{pc}]$ & $1.17 \pm 0.65$ & $0.33 \pm 0.17$ & $9.22 \pm 2.26$\\
$r_{\mathrm{min}} [R_s]$ & $203 \pm 102$ & $58 \pm 27$ & $1598 \pm 1083$\\
$\mathscr{D}[10^{-4} \mathrm{pc}]$& $7.77 \pm 0.38$&$7.77 \pm 0.38$&$1.94 \pm 0.15$\\
$r_{\mathrm{mad}}[ \mathrm{pc}]$ &-&-&$0.010 \pm 0.002$\\
$r_{\mathrm{max}}[\mathrm{pc}]$& $ 2284 \pm 216$& $0.031 \pm 0.005$ & $1.49 \pm 0.79$\\
$\frac{\log(r_{\rm max}/\rm pc)}{\log(r_{\rm min}/\rm pc)}$ &$\frac{+3.35}{-3.93}$ & $\frac{-1.51}{-4.48}$ & $\frac{+0.17}{-3.03}$\\
\hline
$\chi^2_r$ & $1.16$ & $1.15$ & $1.08$\\
$\frac{\Delta \chi^2}{\Delta dof}$ & $38.3$ & $22.3$ & $13.3$ \\
\hline
P$_{\rm null}$ & $9.69 \times 10^{-9}$ & $0.01$ &$2.15 \times 10 ^{-11}$ \\
\hline
\hline
\end{tabular}
{\raggedright \textbf{Note:} The values of the column density, the ionization fraction and the observed redshift of the three absorbers are obtained from the fit with \textsc{xillver} model (see Table \ref{tab:bb_fit}). The upper ($r_{\rm max}$) and the lower ($r_{\rm min}$) limits in the location of the outflows are computed using equation \ref{eq:rmax} and \ref{eq:rmin}, respectively; $r_{\rm mad}$ is computed using the equation \ref{eq:rmad} while $\mathscr{D}$ is the light travel time distance. \par}
\end{table}
The values of the Doppler shifts and the relative outflow velocities for the three absorbers are reported in Tab \ref{tab:doppler_shift}.\\
It is possible to place limits on the radial location of the absorbers, r. From the definition of the ionization parameter $\xi =L_{\rm ion}/nr^2$ \citep{tarter1969} where $L_{\rm ion}$ is the unabsorbed ionizing luminosity emitted by the source between 1\,Ryd and 1000\,Ryd (1 Ryd = 13.6\,eV), $n$ is the number density of the absorbing material and $r$ is the distance from the central source, it is possible to place an upper limit on r, assuming that the thickness of the absorber does not exceed the distance from the SMBH, thus that the absorbers are somewhat compact \citep{Crenshaw2012,Tombesi2013}:
\begin{equation}
r_{\mathrm{max}} = L_{\mathrm{ion}}N_{\rm H}^{-1}\xi^{-1}
\label{eq:rmax}
\end{equation}
with $N_{\rm H}$ and $\xi$ respectively the column density and the ionization fraction of the absorber. We found the ionizing luminosity of the source being  $3.79 \times 10^{43}$ $\rm erg\,s^{-1}$. \\
A lower limit on the radial location of the outflow can be placed estimating the radius in which the observed velocity assumes the value of the escape velocity:
\begin{equation}
r_{\mathrm{min}}=2GM_{\rm BH}v^{-2}
\label{eq:rmin}
\end{equation}
It is possible to place a lower limit on the radial position considering also the light travel time distance: $\mathscr{D} = \Delta t \times c$ during the observing time. Since Wind 1 and Wind 2 does not appear to be variable during the observing time, this limit can be computed using the whole observing time (80\,ks), while, for Wind 3, which shows some level of variability (see Figure \ref{fig:time_cuts_contours}), we used $\Delta t=20$\,ks.\\
An alternative method to compute the radial distance of Wind 3, which show some level of variability during the observation, is to use the median absolute deviation (mad; e.g., \citealp{Serafinelli2019}) of the column density ($N_{\rm H}$), and the outflow velocity ($v_{ \rm w3}$) obtained by fitting the \textit{XMM-Newton} EPIC-pn spectra extrapolated in the 20\,ks intervals (see Table \ref{tab:timecuts}). With the same assumption of \citet{Serafinelli2019}, given the variability of $N_{\rm H}$ and $v_{\rm w3}$, it is possible to compute a typical value of the shell density $\langle n \rangle$:
\begin{equation}
\langle n \rangle =\frac{\Delta N_{\rm H}}{\Delta t \Delta v_{\rm w3}}
\label{eq:madN}
\end{equation}
where $\Delta t=20$\,ks is the time interval in which the observation has been splitted in, $\Delta N_{\rm H}$ and $\Delta v_{\rm w3}$ are the median absolute deviations of the column density and of the outflow velocity, respectively. Then, we computed the distance of the absorber using the following:
\begin{equation}
r_{\mathrm{mad}}=\sqrt{\frac{L_{\mathrm{ion}}}{\langle n \rangle\xi}}
\label{eq:rmad}
\end{equation}
The values of the estimated radial location of the three absorber components in IRAS04416+1215 are reported in Table \ref{tab:doppler_shift}. The lower limits on the distances of the three components are comparable with the distance values extrapolated from the light travel time. Comparing these values with the literature it is possible to see that the Wind 1 and 3 show the typical upper and lower limits of the distance of the WAs for the type\,1 Seyfert galaxies \citep{Tombesi2013} while the Wind 2 in within the range of the average locations of ultra-fast outflows ($\sim 3\times10^{-4}-3\times10^{-2}$ pc, see \citealt{Tombesi2012}). If we only take into account the lower limits, the three winds can be interpreted as three co-spatial absorbers, as the lower limits on the distance are comparable with the dimension of the accretion disc (see the values of $r_{\rm min}$ in unit of Schwarzschild radius in Table \ref{tab:doppler_shift}).  Instead, considering the upper limits and looking at the values of the velocities of the outflows, we can interpret the three wind as multi-phase and multi-scale winds. The Wind 1 can be interpreted as a WA, the Wind 2 as an UFOs and the Wind 3, which shows an intermediate situation between the other two winds, as a so-called entrained ultra-fast outflow (E-UFO, \citealt{Serafinelli2019}).\\
Due to the large discrepancies between the upper and lower limits on the location of the outflows in this dataset, hereafter we will discuss two possible alternative scenarios, one in which the multi-phase winds are co-spatial and the second one in which they are on different scales, based assuming the lower and upper limits, respectively. Looking at the outflow energetics, we computed the mass outflow rate using the following equation from \citet{Crenshaw2012}:
\begin{equation}
\dot{M}_{\rm out}=4\pi r N_{\rm H} \mu m_p C_g v_{\rm w}
\label{eq:mass_outflow_rate}
\end{equation}
where $r$ is the absorber’s radial location (i.e., its distance from the central SMBH), $N_{\rm H}$ is the equivalent hydrogen column density, $\mu$ is the mean atomic mass per proton (= 1.4 for solar abundances), $m_p$ is the mass of the proton, $C_{\rm g}$ is the global covering factor ($ \simeq 0.5$, \citealt{Tombesi2010}) and $v_{\rm w}$ is the radial-velocity centroid.
We computed also the value of the momentum rate (or force) of the outflow, that is rate at which the outflow transports momentum into the environment of the host galaxy, and the mechanical power imparted by expelling mass at a rate $\dot{M}_{\rm out}$ with velocity $v_w$. The momentum rate of the outflow, is given by:
\begin{equation}
\dot{p}=\dot{M}_{\rm out}v_w
\label{eq:momentum_rate}
\end{equation}
The kinetic power of the outflow is obtained by the following relation:
\begin{equation}
\dot{K}=\frac{1}{2}\dot{M}_{\rm out}v_w^2
\label{eq:kinetic_power}
\end{equation}
We computed the mass outflow rate considering all the different values for the radial positions of the outflows and with these values of the mass outflow rate we extrapolate the momentum rates and the kinetic powers. It is interesting to compare the mass outflow rate with the mass accretion rate of the source. Given the dimensionless accretion rate $\dot{\mathscr{M}}\simeq426.58$ \citep{2015ApJ...806...22D}, the mass accretion rate is $\dot{M}_{\bullet}=5.66\,M_{\odot}/yr$. We also compared the momentum rate and the kinetic power of the outflows with the momentum of the radiation of the source, $\dot{p}_{\rm rad}$, and with the outflowing observed bolometric luminosity of the source, which for super-Eddington sources is typically assumed to be $L_{\rm b,out} \sim 100 \times L_{\rm ion}$ \citep{Tombesi2012,2014ApJ...784...77G,Gofford2015},  respectively. All these values are reported in Table \ref{tab:energetic}. The momentum of the radiation of the source is defined as the ratio between the observed luminosity and the velocity of light, thus for IRAS\,04416+1215, it is: $\log(\dot{p}_{\rm rad}/\rm erg\, \rm cm^{-1})=35.04$.\\
\begin{table}
\centering
\caption{Upper and lower limits on the mass outflow rate $\dot{M}$, on the momentum rate of the outflow $P$ and on the kinetic power of the outflow $\dot{K}$ for the absorbers compared with the mass accretion rate ($\dot{M}_{\rm acc}$) the momentum of the radiation ($P_{\rm rad}$) and the observed outflowing bolometric luminosity ($L_{\rm b,out}$), respectively}
\label{tab:energetic}
\begin{tabular}{lccc}
\hline
\hline
Parameter & Wind 1 & Wind 2 & Wind 3\\
\hline
\hline
$\dot{M}_{\rm min}/\dot{M}_{\rm acc}$ & $0.0002$ &$0.002$& $0.002$ \\
$\dot{M}_{\mathscr{D}}/\dot{M}_{\rm acc}$ &$0.001$& $0.042 $& $0.0003$\\
$\dot{M}_{\rm mad}/\dot{M}_{\rm acc}$&-&-&$0.014$\\
$\dot{M}_{\rm max}/\dot{M}_{\rm acc}$ &$3226 $&$1.66$&$2.68$\\
\noalign{\medskip}
$\dot{p}_{\rm min}/\dot{p}_{\rm rad}$ &$0.0011$ &$0.023$& $0.004$ \\
$P_{\mathscr{D}}/\dot{p}_{\rm rad}$ &$0.007$& $0.54$&$0.0008$ \\
$\dot{p}_{\rm mad}/\dot{p}_{\rm rad}$ &-& -&$0.034$  \\
$\dot{p}_{\rm max}/\dot{p}_{\rm rad}$ &$21933$&$21.11$&$6.49$  \\
\noalign{\medskip}
$\dot{K}_{\rm min}/L_{\rm b,out}$&$0.00039$ & $ 0.0015$&$0.00050$  \\
$\dot{K}_{\mathscr{D}}/L_{\rm b,out}$&$0.0002$&$0.035$&$0.00016$\\
$\dot{K}_{\rm mad}/L_{\rm b,out}$&-&-&$0.0004$\\
$\dot{K}_{\rm max}/L_{\rm b,out}$ &$769$ &$1.39$& $0.082$ \\
\hline
\hline
\end{tabular}\\
{\raggedright \textbf{Note:} The mass outflow rate, $\dot{M}$, computed using equation \ref{eq:mass_outflow_rate}, is compared with the mass accretion rate of the source: $\dot{M}_{\rm acc}=5.66\,M_{\odot}/y\rm r$. The momentum rate of the outflow, $P$, and the kinetic power of the outflow, $\dot{K}$, are computed using equation \ref{eq:momentum_rate}  and \ref{eq:kinetic_power}, respectively. The kinetic power of the outflows is compared with the observed outflowing bolometric luminosity of the source: $L_{\rm b,out} \sim 100 \times L_{\rm ion}$ \citep{Tombesi2012,2014ApJ...784...77G,Gofford2015}. We used the same bolometric luminosity to compute the momentum of the radiation of the source, $\log(\dot{p}_{\rm rad}/\rm erg\, \rm cm^{-1})=35.04$, with which the momentum of the outflows is compared.\par}
\end{table}
The typical value of the mass outflow rate for sources accreting below or close to the Eddington limit is $\dot{M}_{\rm out} \gtrsim 5-10\% \dot{M}_{\rm acc}$, for both UFOs and non-UFOs \citep{Tombesi2012}. In this scenario, even for the UFOs with the lowest allowed velocity, the mechanical power is enough to exercise a significant feedback impact on the surrounding environment. Looking at the comparison between mass accretion rate and mass outflow rate for our source (see Table \ref{tab:energetic}), the upper limit on the mass outflow rate for Wind 1 is extremely high, but the values for Wind 2 and 3 are still comparable with the values of quasars and Seyfert galaxies \citep{Tombesi2012}. If we consider the lower limits instead, their values are well below the average. Theoretical works \citep{dimatteo2005, Ostriker2010,King2010,Debuhr2011} showed that, in order to have a significant feedback impact in the environment surrounding an AGN, it is required a minimum ratio between the mechanical power of the outflow and the bolometric luminosity of $\sim 0.5\%$. \citet{Tombesi2012} showed that actually the lower limit of this value for UFOs is $\sim 0.3\%$ and for non-UFOs is $\sim 0.02 \sim 0.8\%$. According with what found for the ratio between the mass outflow rate and the mass accretion rate, looking at the upper limits on $\dot{K}/L_{\rm b,out}$ of our source, i.e. multi-phase and multi-scale X-ray winds, IRAS\,04416+1215 fits well in this scenario in which the outflowing winds can impress a feedback. Indeed, the upper limits on $\dot{K}/L_{\rm b,out}$ is comparable with the kinetic coupling efficiency, defined as the ratio of the kinetic luminosity of outflows to the AGN radiative luminosity ($E_{\rm out}/L_{\rm rad}$), calculated with the feedback model for hyper-Eddington accretion by \citet{2020MNRAS.497..302T}, using the outflow velocities and the $\dot{M}/\dot{M}_{\rm acc}$ values of IRAS\,04416+1215.
Instead, the lower limits are below the minimum value required to generate at least a weak feedback. Considering only the values derived from the lower limits on the distance, i.e. the situation in which the X-ray winds are co-spatial, we would be in a scenario in which the source loses much luminosity due to advection inside the disc, resulting in a much lower efficiency for wind production as most of the radiation remains trapped inside the disc. This deduction is supported also by the results of the lower limits on the ratio between the momentum rate of the outflows and the momentum of the radiation. Outflows accelerated through the continuum radiation pressure are expected to have a $\dot{p}_{\rm out}/\dot{p}_{\rm rad}\sim 1$ \citep{King2015}. The median value of this ratio for UFOs is $\sim 0.96$ after the relativistic correction and $\sim 0.64$ without the relativistic corrections \citep{Luminari2020}. The values we found for the lower limits on $\dot{p}_{\rm out}/\dot{p}_{\rm rad}$ of IRAS\,04416+1215 are again well below the median. Thus, the out-coming luminosity of the source is not enough to accelerate the material to the escape velocity, which is required for a wind to leave the system, suggesting that likely in the scenario of the co-spatial winds the outflows observed in IRAS\,04416+1215 could be accelerated by other mechanisms such as magnetohydrodynamic processes.\\
\begin{figure*}
	\includegraphics[width=0.8\textwidth]{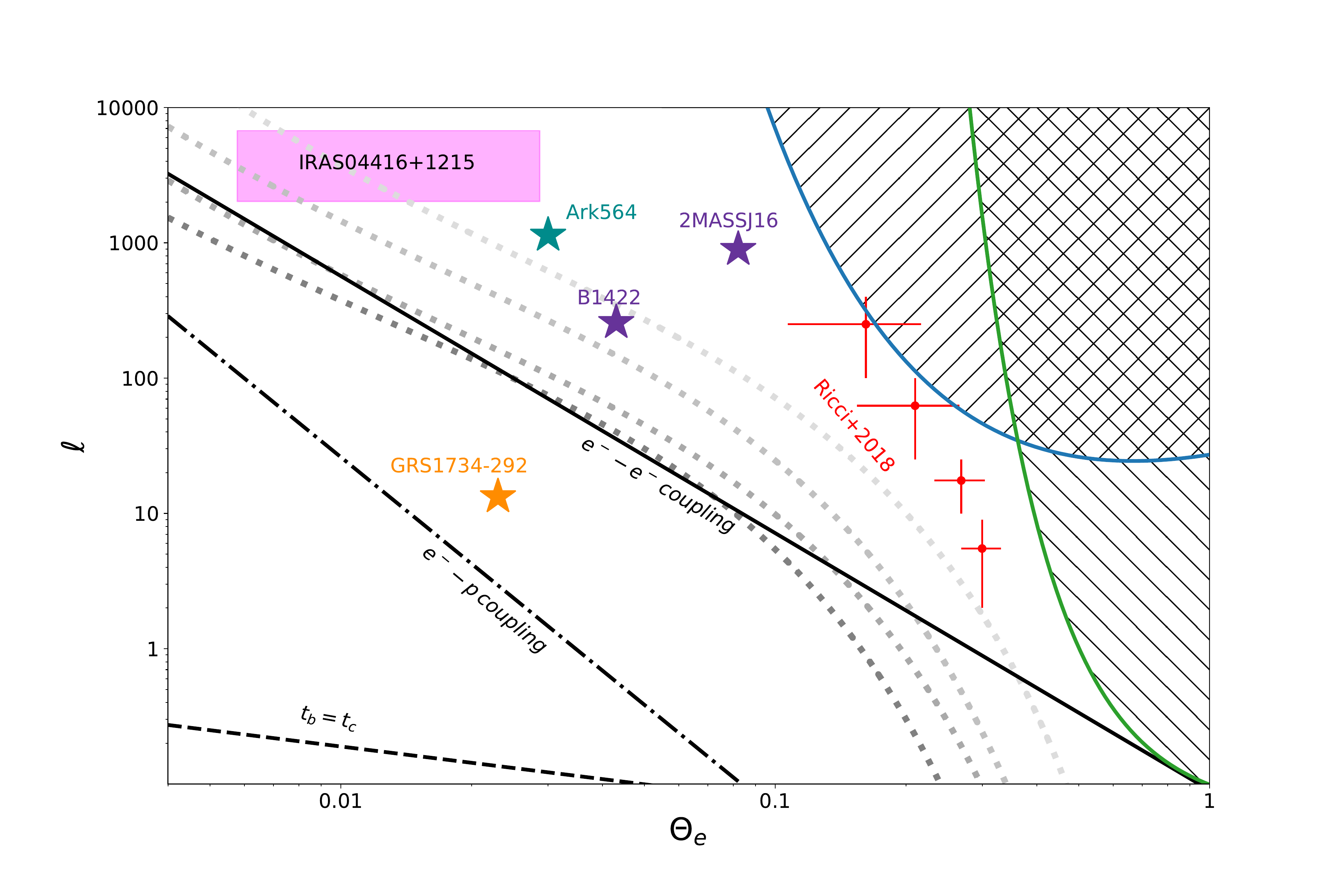}
    \caption{Theoretical compactness-temperature diagram. The black solid line represents the line below which the electron-electron coupling time scale is shorter than the Compton cooling time scale, the black dot-dashed line represents the line below which the electron-proton coupling time scale is shorter than the Compton cooling time scale while the black dashed line represents the line below which the dominant process is the Bremsstrahlung. The blue and green solid curves are the pair run-away lines respectively for a disc-like \citep{Svensson1984} or a spherical corona \citep{Stern1995}. The dotted gray curves are the $\Theta_e - \ell$ distribution for different values of the fraction of thermal, $\ell_{\rm th}$, and non-thermal, $\ell_{\rm nth}$, electrons, for a given ratio between the injection of soft photons ($\ell_{\rm s}$) and the total heating ($\ell_{\rm h}=\ell_{\rm th} + \ell_{\rm nth}$) of $\ell_{\rm s}/\ell_{\rm h}=0.1 $ [from darker to lighter: $\ell_{\rm nth}/\ell_{\rm th}=(0.33; 0.29; 0.23; 0.17)$, see \citealt{Fabian2017}]. The magenta region indicate the position of IRAS\,04416+1215 considering the temperature parameter extrapolated using both the coronal temperature value from the fit with \textsc{xillverCp} table ($kT_e = 2.95$\,keV) and from the relation with the cut-off energy ($kT_e = 10$\,keV) and considering a compactness parameter assuming a radius of 10$R_g$ (following \citealt{Fabian2015}) and 3$R_g$ (following \citealt{2014ApJ...787...83M}). The dark cyan and the orange stars indicate the position of Ark\,564 \citep{Kara_2017} and GRS1734-292 \citep{Tortosa_2016}. The purple stars indicate the position of two luminous high-redshift quasars (2MASSJ1614346+470420  and B1422+231, \citealt{Lanzuisi2019}). The red points represent the median values of the of compactness-temperature diagram for the BASS hard X-ray selected sample from \citet{Ricci_2018}.}
    \label{fig:compactness-temperature_diagram}
\end{figure*}
\subsection{The extremely low coronal temperature of IRAS\,04416+1215}
We estimated the coronal temperature by fitting the spectra with the \textsc{xillverCp} and \textsc{relxillCp} models (see Section \S \ref{sect:comptonization}), and we estimated the optical depth using the relation from \citet{Pozdnyakov1997}:
\begin{equation}
\Gamma \sim 1+ \frac{\left[ \frac{2}{\Theta_e+3} -\log(\tau)\right]}{\log(12\Theta_e^2+25\Theta_e)}
\label{eq:gamma}
\end{equation}
where $\Gamma$ is the photon index of the spectrum between 2 and 10\,keV. The dependence from
 the optical depth is in the relativistic $y$ parameter is:
\begin{equation}
y=4(\Theta_e+4\Theta_e^2)\tau(\tau+1)
\label{eq:comptonparam}
\end{equation}
where $\Theta_e$ is the is the electron temperature normalized to the electron rest energy:
\begin{equation}
\Theta_e=\frac{kT_e}{m_ec^2}
\label{eq:theta}
\end{equation}
Using the coronal temperature value from the fit with \textsc{xillverCp} table ($kT_e=3.0 \pm 0.3$) we extrapolated $\Theta_e = 0.005$ and we found an optical depth $\tau=10.56 \pm 2.11$. The electron temperature of the corona is expected to be related to the spectral cut-off energy, $E_{\rm cut} \sim 2-3 \times kT_e$ \citep{Petrucci2000,Petrucci2001}, so we computed also the value of the temperature parameter and the optical depth using the coronal temperature extrapolated from the best fit with \textsc{xillver} model (see Table \ref{tab:bb_fit}) and we obtained $\Theta_e = 0.028$ and $\tau=2.48 \pm 0.49$. While these values may seem extreme, they are in agreement with previous studies \citep{Pounds1995,Brandt1997,Shemmer2006,Shemmer2008,Risaliti2009,Brightman2013,Kawamuro2016, Ricci_2018}, which suggested that, for low-mass black holes accreting close to the Eddington limits, a steep spectrum and a low value of the AGN coronal temperature is expected. So far, the existence of of relatively cool ($kT_e \lesssim10$\,keV) and optically thick ($\tau\gtrsim3$) coronae is confirmed for ULXs \citep{Yoshida2013, Shidatsu2017,2020MNRAS.494.6012W,Walton2021,middleton2021nustar}, and black hole X-ray binaries (XRBs; e.g. GRS\,1915+105, see \citealt{Vierdayanti2010,koljonen2021almanicer}), with super-Eddington accretion flow.\\
From the theoretical point of view, a recent work from \citet{kawanaka2021determines} shows that considering an outflowing corona formed in a black hole accretion flow above the Eddington value, fed by the radiation pressure-driven wind from an underlying disc, and heated by the reconnection of magnetic loops emerging from the disc, the coronal temperature can be a few tens of keV, and it would be cooler for higher mass accretion rate. Moreover, in contrast with the case of sub-Eddington accretion, where the coronal optical depth is around unity, the peak value of the optical depth of outflowing corona could be $1\lesssim \tau \lesssim 10$.\\
For completeness, we compared  our results of IRAS\,04416+1215 with the results of \citet{Fabian2015} and \citet{Ricci_2018} to see the location of  IRAS\,04416+1215 in the compactness-temperature ($\Theta_e-\ell$; \citealt{Fabian2015} and references therein), where $\Theta_e$ is the aforementioned electron temperature normalized to the electron rest energy, and $\ell$ is the dimensionless compactness parameter \citep{Fabian2015}:
\begin{equation}
\ell=\frac{L}{R}\frac{\sigma_T}{m_ec^3}
\label{eq:ell}
\end{equation}
where L is the luminosity, R is the radius of the corona and $\sigma_T$ is the Thompson cross section.\\
To compute the compactness parameter of IRAS\,04416+1215 we adopted the luminosity of the power-law component extrapolated to the 0.1--200\,keV band and we assume two limit values for the radius: a value of 10 gravitational radii R$_g$ (for standard sources, following \citealt{Fabian2015}) and 3 R$_g$ (for super-Eddington sources, following \citealt{2014ApJ...787...83M}, obtaining respectively $\ell=2024$ and $\ell=6748$.\\
The compactness temperature diagram is very useful to understand the various physical properties of a physical finite, thermal plasma. The dominant radiation process in a plasma will be the one with the shortest cooling time. In the AGN hot corona the most significant processes are the Bremsstrahlung, the inverse Compton scattering and the pair production. Comparing the cooling times of the different processes it is possible to define some regions in which one process is dominant over the other. Comptonization dominates at high compactness ($\ell > 3\alpha_f\Theta_e^{-1/2}$, where $\alpha_f$ is the fine-structure constant), when $3\,\alpha_f \Theta_e < \ell < 0.04\,\Theta_e^{-3/2}$ the dominant effect is the electron-proton coupling while for $0.04\,\Theta_e^{-3/2}< \ell<80\,\Theta_e^{-3/2}$ the electron-electron coupling becomes relevant \citep{1994ApJS...92..555F}. Beyond a certain regime the pair production becomes a runaway process. In the $\theta_e -\ell$ plane this regime is identified by the, so-called, pair runaway lines. The position of these lines depends on the shape of the source and on the radiation mechanism. \citet{Stern1995} computed the pair balance curve for a slab corona (green line in Figure \ref{fig:compactness-temperature_diagram}). \citet{Svensson1984} estimated that the pair balance for an isolated cloud occurs when $\ell \sim 10\,\Theta_e^{5/2}\,e^{1/\Theta_e}$ (blue line in Figure \ref{fig:compactness-temperature_diagram}).\\
 Considering the value of the coronal temperature obtained for the case of standard reflection and the comptonization model \textsc{xillverCp}, IRAS\,04416+1215 is located on the edge of the e$^--e^-$ coupling line, in which the e$^--e^-$ coupling time scale is longer than the Compton cooling time scale, well below the forbidden region bounded by the pair run-away line for both slab and spherical corona, in which the pair production exceeds the annihilation and the pair production becomes a run-away process. In the case of the coronal temperature extrapolated from the cut-off energy parameter, using the relation from \citet{Petrucci2000,Petrucci2001}, IRAS\,04416+1215 is located further away from the e$^--e^-$ coupling line, but still far from the pair run-away region (see Figure \ref{fig:compactness-temperature_diagram}).
In the works of \citet{Fabian2015}, \citet{2014ApJ...787...83M} and \citet{Ricci_2018}, most of the sources are located close to the pair run-away line suggesting that the outgoing AGN spectral shape is controlled by pair production and annihilation. In fact in a compact corona (i.e., $\ell >1$) the photon density is high and if the photons are energetic enough, photon-photon collisions can lead to pair production. If the temperature reaches values $\sim 1\, \rm MeV$, pair production becomes significant, consuming energy and limiting the rise in temperature. The location of IRAS\,04416+1215 within the $\Theta_e-\ell$ diagram suggests that the corona of IRAS\,04416+1215 is mainly composed of a thermal plasma which is not pair-dominated. But in such a hyper-Eddington source we expect the corona being highly magnetized and powered by dissipation of magnetic energy. Thus the corona could be an hybrid plasma, composed of both thermal and non-thermal electrons (e.g., \citealp{1993ApJ...414L..93Z,Fabian2017}). In this scenario the heating and cooling are so intense that the energetic particles may not have time to thermalize before inverse Compton cooling reduces their energy. Therefore if a fraction of electrons that do not follow the thermal distribution exceeds the temperature of $\sim 1\, \rm MeV$ they can emit hard photons which, colliding, can create electron-positron pairs and, if the pairs are energetic enough, i.e. $\gtrsim 2m_ec^2$, a run-away situation can occur because more pairs are produced. Before annihilating, the cooled pairs soak up energy, and the mean energy per particle, and the temperature of the thermal electrons, decreases. Then, the primary continuum emission would appear to originate in a coronal plasma with low-temperature which is not pair-dominated. Recently, the observations of GRS\,1734-292 \citep{Tortosa_2016} and Ark\,564 \citep{Kara_2017} showed these sources have a very low coronal temperatures, but not as low as IRAS\,04416+1215 (see Figure \ref{fig:compactness-temperature_diagram}). Ark\,564 is a high-Eddington NLS1, so its corona could also have a hybrid plasma of thermal and non-thermal electrons, in agreement with the scenario of the extremely low coronal temperature of IRAS\,04416+1215. Instead, GRS\,1734-292 is accreting at a few percent of the Eddington limit, so that the effectiveness of the cooling mechanism cannot be related to a particularly strong radiation field. Instead its low coronal temperature could be associated to its high optical depth value, so that the disc seeds photons would undergo a large number of scatterings, which would reduce the electron temperature.

\section{Conclusions}
\label{sect:conclusion}

Super-Eddington accretion is extremely important to explain the fast growth of the first supermassive black holes, as well as to interpret the properties of tidal disruption events and ultra-luminous X-ray binaries. 
Here we have presented the detailed broad-band analysis of the first simultaneous \textit{XMM-Newton} and \textit{NuSTAR} observations of IRAS\,04416+1215, part of a dedicated campaign to study eight super-Eddington AGN from the Super-Eddington Accreting Massive Black Holes (SEAMBHs) sample \citep{Du2014,Wang2014a,2015ApJ...806...22D}. The SEAMBHs objects have the black hole masses estimated by reverberation mapping, which allows to accurately constrain the Eddington ratio. IRAS\,04416+1215 is the most peculiar object of our campaign and it has the highest Eddington ratio ($\lambda_{\rm Edd}\simeq 472$) among the sources of our sample and one of the highest in the local Universe.

In summary the results of our analysis are the following:
\begin{list}{-}{\setlength{\itemsep}{0.cm}}
    \item IRAS\,04416+1215 shows a peculiar light curve in the \textit{XMM-Newton} energy band (i.e., 0.2--10\,keV). There is a rapid increase followed by a decrease of the count rate by a factor two in less than two hours during the observation. However, this variability does not affect the overall spectral shape. The X-ray variability spectrum of IRAS\,04416+1216 shows a different trend respect to the median of the Seyfert\,1 sample from BASS with almost the same black hole mass as IRAS\,04416+1215 (i.e., $\sim 3\times10^6 -1\times10^7$M$_{\odot}$) with a higher variability in the soft band, suggesting that the most variable components are the ones found at low energy, such as the soft excess and/or ionized absorption \citep{1997ApJ...476...70N,2012yCat..35420083P,10.1093/mnras/stu2618}.
    \item We carried out a detailed spectral analysis of the broad-band data obtained from the simultaneous observation of the source by \textit{XMM-Newton} and \textit{NuSTAR} using a large variety of spectral models. The spectral shape of the source appears to be very similar to the standard NLS1 galaxy's spectra. The best-fitting model is composed of a soft-excess, three ionized outflows, neutral absorption and a reflection component, which was modelled with \textsc{xillver} in \textsc{xspec}. We found that the reprocessed radiation is dominating ($R_{\rm refl} > 8.01$) over the primary continuum, which shows a slope of $\Gamma=1.77$, as in the case of other sources with high accretion rate \citep{2009Natur.459..540F,2014ApJ...795..147R}.
    \item  The source shows the presence of a multi-phase absorption structure composed of three phases. 
    From the analysis of the location and the energetic of these X-ray winds, two main scenarios arise: in the first one the three winds can be considered as co-spatial absorbers, in the second one we have Wind 1 and 2 showing the parameters consistent with being a WA and a UFO, respectively, while Wind 3 shows the features of an E-UFO. In the former scenario, the disc emitted luminosity does not have enough power to accelerate the winds because, due to the extremely high accretion rate, much of the luminosity is advected inside the disc and the winds receive a relatively low acceleration from radiation pressure alone. Thus, suggesting that the outflows observed in IRAS\,04416+1215 could be accelerated by magnetohydrodynamic processes. In the latter scenario, instead, the estimated power suggests that the three multi-scale winds produce an important feedback on the environment surrounding the AGN.
    \item IRAS\,04416+1215 has a very well constrained value of the coronal temperature, one of the lowest coronal temperature measured so far by \textit{NuSTAR}. Depending on the modelling we found a coronal temperature between $\sim 3$\,keV and $\sim 22$\,keV. The location of IRAS\,04416+1215 in the compactness-temperature diagram suggests that, unlike most of the AGN, its X-ray emission is not created in a pair-dominated thermal plasma. However, considering a scenario in which the corona is made of an hybrid plasma, composed of both thermal and non-thermal electrons, the non-thermal electrons that exceed the temperature of $\sim 1$\,MeV can emit photons which give rise to a pair run-away situation. In this case, the primary continuum emission would seem to originate in a thermal coronal plasma with low-temperature which is not pair-dominated. A possible explanation for this very low value of the coronal temperature of IRAS\,04416+1215 can be found considering that, in Super-Eddington accretions flows, the comptonizing corona could originate from radiation-pressure driven optically thick ($\tau \gtrsim 3$) outflows which act like a corona above the disc at relatively low temperatures ($kT_e \lesssim 10$\,keV, see \citealt{kawanaka2021determines}).
\end{list}
In our spectral analysis of IRAS\,04416+1215 we tried different models which appeared to be statistically as good as the best-fitting model (i.e., \textsc{xillver}, see Section \S \ref{sect:bbanalysis}), even if they showed some weaknesses. However, in the best-fitting model we could not fully constrain the reflection fraction, and a we could not discriminate between two different scenarios on the origin of the ionized outflows. To better understand the properties of this source, a further simultaneous \textit{NuSTAR} (170 ks) and \textit{XMM-Newton} (40ks) observation of IRAS\,04416+1215 with longer exposures will be performed during the \textit{NuSTAR} cycle 7.\\
Moreover, in a forthcoming paper (Tortosa et al. in prep.) we will present the analysis of all sources of our sample, which will provide the first systematic broad-band X-ray study of extreme accretion onto SMBHs with simultaneous \textit{NuSTAR} and \textit{XMM-Newton} observations.


\section*{Acknowledgments}
AT acknowledge the support from FONDECYT Postdoctorado for the project n. 3190213. CR acknowledges support from the Fondecyt Iniciacion grant 11190831. LCH was supported by the National Science Foundation of China (11721303, 11991052) and the National Key R\&D Program of China (2016YFA0400702). JMW acknowledges financial support by the National Science Foundation of China (NSFC) through grants NSFC-11991054 and -11833008, and the National Key R\&D Program of China (2016YFA0400701). PD acknowledges financial support from NSFC grants NSFC-12022301, -11873048, and -11991051, and from the Strategic Priority Research Program of the CAS (XDB23010400). This work is based on observations obtained with the ESA science mission \textit{XMM-Newton}, with instruments and contributions directly funded by ESA Member States and the USA (NASA), and the \textit{NuSTAR} mission, a project led by the California Institute of Technology, managed by the Jet Propulsion Laboratory and funded by NASA. This work is financially supported by the National Science Foundation of China (11721303, 11991052, 11950410493, 12073003), the National Key R\&D Program of China (2016YFA0400702). This research has made use of the \textit{NuSTAR} Data Analysis Software (NuSTARDAS) jointly developed by the ASI Space Science Data Center (SSDC, Italy) and the California Institute of Technology (Caltech, USA). The authors thank Javier García for kindly providing the beta-version of the \textsc{xillverDCp} and \textsc{relxillDCp} tables and the referee for the useful suggestions which helped in improving the manuscript.\\

\section*{Data Availability}

 All the data utilized in this paper are publicly available in the \textit{XMM-Newton} and \textit{NuSTAR} archives. More details of the observations are listed in Table \ref{tab:observations_table}.


\bibliographystyle{mnras}
\bibliography{bibliography} 

\appendix
\section{Detailed fitting procedure}
\label{app:fitting}
The fit procedure was carried out at first keeping  all the parameters fixed to the values we found while fitting the 0.35--3\,keV and the 3--24\,keV spectra, then allowing them to vary one by one. For the primary continuum and the reprocessed radiation, following what we have done for the hard X-ray band, we tested first the reflection with the reflection model \textsc{xillver}. After this process we obtained a fairly good fit, with a $\chi^2$ of 854 for 788 dof. The primary continuum showed a power-law ($\Gamma = 1.77$) with a cut-off at $\sim 44$\,keV. The best-fitting parameters are listed in Table \ref{tab:bb_fit}, the spectra, the fitting model and the residuals are shown in rigth panels of Figure \ref{fig:baseline}. Since for AGN accreting at high accretion rate the disc is expected to be much denser than the standard assumption (see Fig. 1 of \citealt{Garcia2016}), we replaced \textsc{xillver} with \textsc{xillverD}, which is the same model as the standard reflection but it allows a higher density for the accretion disc, ranging from the standard case of $\log(n/\rm cm^{-3})$=15 (i.e. \textsc{xillver}) up to $\log(n/\rm cm^{-3})$=19 \citep{Garcia2016}.
The fit is slightly worse compared to the previous case ($\chi^2$=862 for 788 dof). We found a disc density value of $\log(n/\rm cm^{-3})$=$16.81^{+0.65}_{-2.42}$. The photon index appeared to be even steeper ($\Gamma=2.04\pm0.09$), but it should be stressed that in this model the high-energy cut-off is fixed at 300\,keV. The following step of the analysis was to replace the \textsc{xillver} model with the \textsc{relxill} version [1.4.3] model \citep{Garcia2014, Dauser2014}, in order to test for the presence of relativistic reflection. We fixed the inclination angle to a value of $30^{\circ}$. The iron abundance, the ionization parameter and reflection fraction were allowed to vary together with the photon index and the cut-off energy. We allowed also the spin parameter to vary while the disc inner radius was defined by the spin (i.e., R$_{\rm in} \equiv \rm{R}_{\rm isco}$, see \citealt{1969ApJ...156..943T}). This model provided a good fit ($\chi^2$ of 847 for 787 dof). The photon index of the primary power-law has the same value as for the fit with \textsc{xillver} model, while the cut-off energy almost doubled ($E_{\rm cut}\sim 80$\,keV, see Table \ref{tab:bb_fit}).
\begin{figure}
\centering
	\includegraphics[width=\columnwidth]{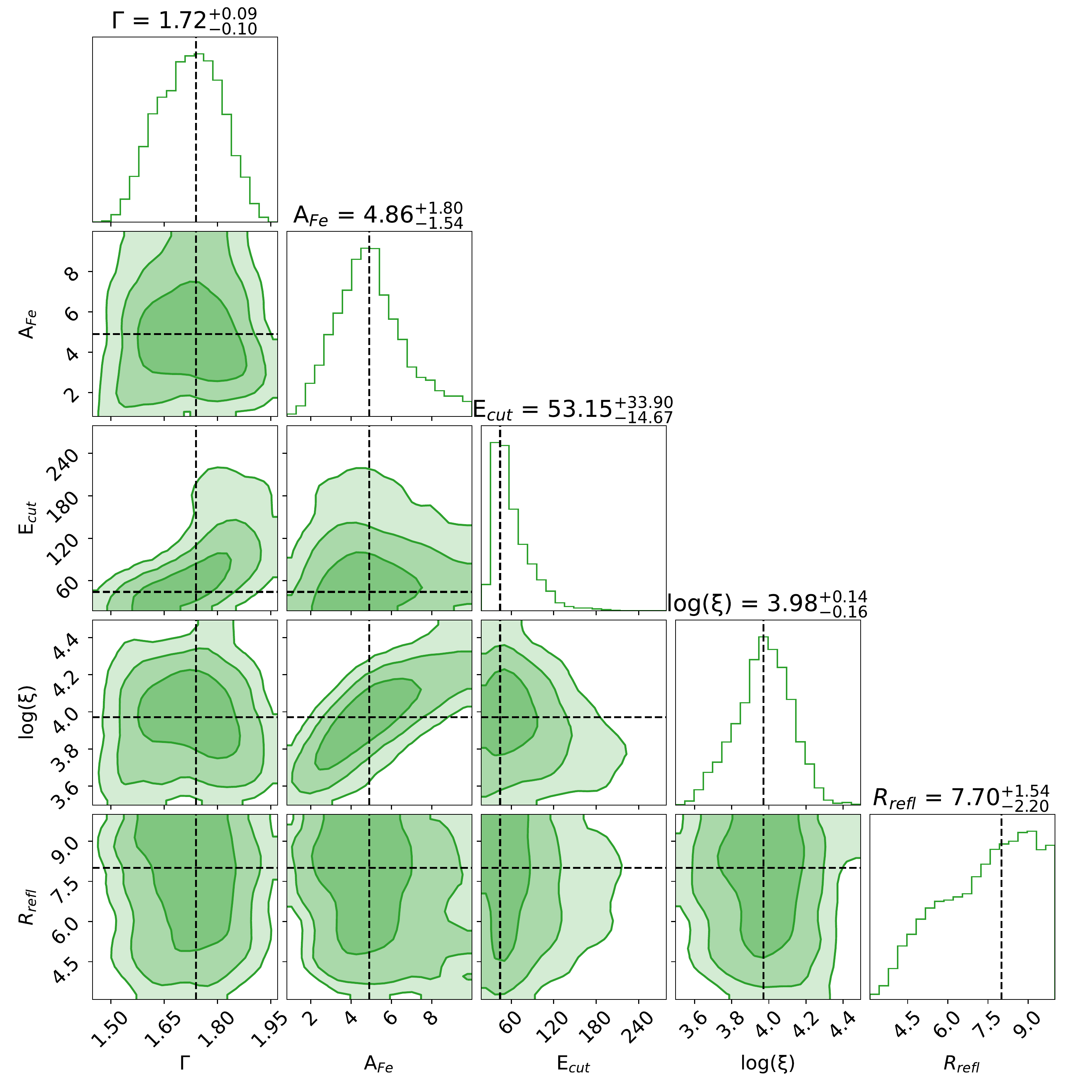}
    \caption{68\%, 90\% and 99\% contour plots resulting from the MCMC analysis of the \textsc{xillver} model applied to the broad-band \textit{XMM-Newton} and \textit{NuSTAR} 0.35--24\,keV spectra of IRAS\,04416+1215. We show the outputs for photon index ($\Gamma$), iron abundance ($A_{\rm Fe}$), cut-off energy ($E_{\mathrm{cut}}$[keV]), ionization parameter ($\log(\xi /\rm erg\,\rm cm\,\rm s^{-1}$)) and reflection fraction ($R_{\rm refl}$). }
    \label{fig:triangular}
\end{figure}
However, the black hole spin parameter was not constrained, we obtained just a lower limit ($a > -0.32$ where negative values mean the accretion disc is counter-rotating with respect to the black hole). Next we replaced the standard relativistic reflection with the \textsc{relxillD} model (i.e., the relativistic reflection including the case of the high density accretion disc, \citealt{Garcia2016}). The fit slightly improved ($\chi^2$=844 for 787 dof), and we found a disc density of $\log(n/\rm cm^{-3}$)=$18.01^{+0.26}_{-2.73}$ and a lower limit for the spin.\\
In all the models tested we found an iron overabundance and a high value of the reflection fraction (see Table \ref{tab:bb_fit}), confirming the result obtained by fitting only the 3--24\,keV energy range.\\
We checked the value of the disc inclination angle allowing this parameter to vary in all the fitting models we tested. We obtained an upper limit ranging from $\sim 25^{\circ}$ to $\sim45^{\circ}$, depending on the model. Since the statistical significance of the fits was the same with respect to the fit with the inclination parameter fixed to the default value of $30^{\circ}$ for each model and since the other parameters did not changed, we preferred to keep it fixed to the default value.\\
For completeness, we also tested the hypothesis of the presence of two reflection component, one distant neutral reflection from the torus, modelled with \textsc{xillver} with the ionization fraction parameter pegged to zero, and the  disc relativistic reflection, modelled with \textsc{relxill}. The fit gave a $\chi^2$ of 888 for 808 dof. We obtained a cutoff value well constrained, $E_c = 10^{+7}_{-5}$\,keV, and a photon index of $\Gamma=1.46\pm 0.03$. The reflection fraction of the disc reflection component is consistent with zero and the spin value is unconstrained with an upper limit of $\sim 0.85$. The distant neutral reflection component showed a reflection fraction $R_{\rm refl}=0.23 \pm 0.12$. These values lose physical significance in the context of such a rapidly accreting source, for which a steep spectral shape, reflection dominated is expected. Besides, the statistical significance of the fit is worse with respect to models considering only one reflection component. For these reasons we exclude the hypotesis of this scenario.\\
We searched for possible degeneracies between the fitting parameters, performing Monte Carlo Markov Chain (MCMC) using \textsc{xspec-emcee} code by Jeremy Sanders\footnote{\url{https://github.com/jeremysanders/xspec_emcee}}. This is an implementation of the emcee code \citep{ForemanMackey2013}, to analyze X-ray spectra in \textsc{xspec}. We used 50 walkers with 10,000 iterations each, burning the first 1,000. The walkers started at the best fit values found in \textsc{xspec}, following a Gaussian distribution in each parameter, with the standard deviation set to the delta value of that parameter. The MCMC results of the fitting models applied to the broad-band data are shown in Figure \ref{fig:triangular} for the most relevant parameters: photon index ($\Gamma$), iron abundance ($A_{\rm Fe}$), cut-off energy ($E_{\mathrm{cut}}$/keV), ionization parameter ($\log(\xi /\rm erg\,\rm cm\,\rm s^{-1}$)) and reflection fraction ($R_{\rm refl}$).\\
As stated previously, being IRAS\,04416+1215 a Super-Eddington accreting source, we expected for this source a black hole accretion disc spectra different from the one resulting from standard sub-Eddington accreting sources. Thus, as a last test, we fitted IRAS\,04416+1215 spectra using the AGN super-Eddington accretion model \textsc{agnslim} which is based on the slim disc emissivity \citep{1988ApJ...332..646A,2000PASJ...52..133W,2011arXiv1108.0396S}. This test ended out to be unsuccessfully since the fitting parameters did not converge to any physically acceptable values. Therefore, we discarded this model.
\section{Tables of the fitting parameters.}
\label{app:tables}
\begin{table}
\caption{Winds parameters from the best fits of the 4 \textit{XMM-Newton} EPIC-pn spectra extrapolated in the 20\,ks time intervals. Errors are at 90\% confidence levels.}
\label{tab:timecuts}
\begin{tabular}{llcccc}
\hline
\hline
Comp. & Param. & $\Delta t_1$&$\Delta t_2$&$\Delta t_3$&$\Delta t_4$\\
\hline
\hline
Wind 1 & $N_{H}$ &$0.55^{+0.13}_{-0.03}$ & $0.56^{+0.05}_{-0.02}$&$0.53^{+0.05}_{-0.02}$&$0.53\pm 0.09$\\
Wind 1  & $\log(\xi)$ & $<0.23 $& $<0.11$ & $<0.18$ & $<0.18$\\
Wind 1  & z  & $>0.08$ & $0.02^{+0.04}_{-0.06}$ &$0.04\pm 0.06$ & $0.043^{+0.02}_{-0.07}$\\
\hline
Wind 2 & $N_{H}$ & $1.09^{+1.18}_{-0.93}$ & $1.82^{+0.78}_{-0.99}$ & $2.43^{+1.24}_{-.39}$ & $1.67\pm 1.05$ \\
Wind 2 & $\log(\xi)$ & $2.59^{+2.83}_{-0.19}$& $2.70 \pm 0.19$ & $2.66^{+0.13}_{-0.07}$ & $2.66^{+0.29}_{-0.27}$ \\
Wind 2  & z  & $-0.04^{+0.09}_{-0.21}$ & $0.06\pm0.03$ & $0.07\pm 0.01$ & $0.08^{+0.01}_{-0.11}$\\
\hline
Wind 3 & $N_{H}$ & $5.10^{+1.77}_{-2.15}$& $>100$ & $1.92^{+2.03}_{-1.26}$ & $0.13^{+1.47}_{-1.02}$\\
Wind 3  & $\log(\xi)$ & $2.02 ^{+0.63}_{-0.31}$ & $4.09^{+1.44}_{-0.32}$ & $3.10^{+1.20}_{-0.27}$ & $1.90^{+3.79}_{-0.48}$ \\
Wind 3 & z & $0.04^{+0.01}_{-0.07}$ & $-0.07 \pm 0.03 $& $-0.09\pm 0.08$ & $ -0.25^{+0.16}_{-0.04}$\\
\hline
\multicolumn{2}{l}{$\frac{\chi^2}{\rm dof} = \chi^2_r$} &\multicolumn{4}{l}{$\frac{601}{570}=1.05$} \\
\hline
\hline
\end{tabular}
{\raggedright \textbf{Note:} The values of the column density are given in unit of $(10^{22}\rm cm^{-2})$. The ionization fraction is given in unit of $\log(\xi  / \rm erg\,\rm s^{-1} \rm cm)$.\par}
\end{table}

\begin{center}

\begin{landscape}
\begin{table}
\centering
\caption{ Fitting parameters for the X-ray broad-band \textit{XMM-Newton} plus \textit{NuSTAR} spectra of IRAS\,04416+1215 using different \textsc{xillver} and \textsc{relxill} tables for the primary continuum and the reprocessed emission. Errors are at 90\% confidence levels. If a model is calculated by default for a fixed value of a parameter, the value is shown in boldface. }
\label{tab:bb_fit}
\begin{tabular}{l|l|cccccccc|}
\hline
\hline
Component & Parameter & Xillver & Relxill & XillverD & RelxillD & XillverCp & RelxillCp & XillverDCp & RelxillDCp\\
\hline
\hline
Wind1 & $N_{H} (10^{21}\rm cm^{-2})$  & $8.49_{-1.44}^{+1.18}$ &$5.28_{-0.31}^{+1.20}$& $5.17_{-0.46}^{+0.75}$&$5.47_{-0.43}^{+0.67}$& $5.90 \pm 0.40$ & $5.86 \pm 0.50$  &$9.41 \pm 4.01$ & $2.30 \pm 1.01$\\
Wind1  & $\log(\xi  / \rm erg\,\rm s^{-1} \rm cm)$ &  $0.127_{-0.05}^{+0.03}$ & $<0.078$&  $0.0\pm0.050$&$0.003_{-0.001}^{+0.064}$& $<0.03$ & $<0.05$ & $1.66 \pm 0.10 $ & $0.22 \pm 0.15$ \\
Wind1  & z   & $0.015_{-0.021}^{+0.016}$  & $0.008 \pm 0.018$& $0.026_{-0.023}^{+0.014}$ & $-0.001_{-0.012}^{+0.026}$& $0.005 \pm0.012$ & $0.003 \pm 0.02$ & $>0.078$& $>0.068$ \\
\hline
Wind2 & $N_{H} (10^{22} \rm cm^{-2})$ & $2.81_{-3.13}^{+1.24}$ & $10.90_{-0.77}^{+1.45}$ & $1.27\pm1.04$ & $5.66\pm4.85$& $10.0^{+2.56}_{-8.75}$ & $0.65\pm 5.41$ &$1.77 \pm 0.85$  & $0.47 \pm 0.25$ \\
Wind2& $\log(\xi  / \rm erg\,\rm s^{-1} \rm cm)$ & $2.68_{-0.07}^{+0.08}$ & $3.57_{-0.25}^{+0.86}$& $3.26_{-0.34}^{+2.57}$&$3.58_{-0.19}^{+2.19}$& $3.61^{+0.89}_{-0.55}$ & $3.01^{+0.59}_{-0.47}$ & $2.69 \pm 0.08$ & $2.24 \pm 0.12$\\
Wind2& z &$-0.046_{-0.07}^{+0.01}$  & $-0.041_{-0.223}^{+0.095}$& $-0.116 \pm 0.013$ &$-0.046_{-0.186}^{+0.064}$& $-0.04 \pm 0.01$ & $-0.04 \pm 0.02$ & $-0.135 \pm 0.007$ & $-0.014 \pm 0.02$ \\
\hline
Wind 3 & $N_{H} (10^{22} \rm cm^{-2})$ &  $2.16_{-0.67}^{+1.01}$ & $1.89_{-1.15}^{+0.53}$ & $1.46_{-0.76}^{+0.63}$ & $1.29_{-0.71}^{+0.76}$& $4.67^{+1.44}_{-1.16}$ & $2.50^{+1.27}_{-1.18}$ & $1.38 \pm 0.73 $ &  $4.01 \pm 0.58$\\
Wind 3 & $\log(\xi  / \rm erg\,\rm s^{-1} \rm cm)$ &  $2.41_{-0.31}^{+0.13}$ &$2.64_{-0.24}^{+0.08}$ & $2.63_{-0.15}^{+0.08}$&$2.62_{-0.14}^{+0.09}$& $2.72 \pm 0.04$ & $2.64^{+0.06}_{-0.04}$&  $2.34 \pm 0.15$ &  $2.70 \pm 0.12$ \\
Wind 3 & z & $0.062\pm0.009$ & $0.064 \pm 0.020$ & $0.048_{-0.026}^{+0.013}$ & $0.062_{-0.022}^{+0.016}$ &$0.063 \pm 0.008$ & $0.06 \pm 0.01$ & $-0.027 \pm 0.005$ & $-0.051 \pm 0.015$\\
\hline
 Neutral Absorption  & $N_{H}(10^{22} \rm cm^{-2})$ & $0.106_{-0.036}^{+0.053}$ &  $0.099_{-0.025}^{+0.063}$&$0.148_{-0.032}^{+0.063}$ &$0.161_{-0.034}^{+0.050}$& $0.141 \pm 0.05$ & $0.158 \pm 0.037$ & $0.072 \pm 0.030$  & $0.297\pm 0.06$\\
\hline
Black Body & $kT_{ \rm BB}$(keV) & $0.094 \pm 0.005$ & $0.090_{-0.006}^{+0.004}$ & $0.099_{-0.004}^{+0.002}$ &$0.085_{-0.005}^{+0.004}$& $0.093 \pm 0.005$ & $0.089 \pm 0.006$ &$0.106 \pm 0.005$ & $0.083 \pm 0.004$\\
\hline
Ionized reflection continuum & $\mathit{a}$ & \textbf{0} & $>-0.32$& \textbf{0} & $>-0.59$& \textbf{0} & $<0.35 $ & \textbf{0} & $>0.71$ \\
Ionized reflection continuum& $\Gamma$ & $1.77_{-0.09}^{+0.17}$ & $1.77_{-0.09}^{+0.12}$ & $2.04 \pm 0.09$ &$1.75_{-0.13}^{+0.15}$& $1.94\pm0.08$ & $1.84 \pm 0.11$ & $1.88 \pm 0.04$ & $1.69 \pm 0.06$ \\
Ionized reflection continuum  & $\log(\xi /erg s^{-1}cm)$ & $3.97_{-0.30}^{+0.25}$ & $3.39_{-0.15}^{+0.18}$&$3.69_{-0.13}^{+0.44}$ &$3.39_{-0.16}^{+0.12}$& $3.62\pm0.11$ & $1.42^{+0.016}_{-0.08}$ & $4.23 \pm 0.11$ & $3.76^{+0.29}_{-0.18}$\\
Ionized reflection continuum& A $_{\mathrm{Fe}}$  & $5.01_{-2.63}^{+2.98}$ & $> 7.61$&$2.50_{-0.74}^{+1.64}$ &$>8.14$& \textbf{1} & \textbf{1} & $5.00^{+3.02}_{-0.95}$ & $>3.73$ \\
Ionized reflection continuum &  $E_{\mathrm{cut}}$(keV) & $44.19_{-17.45}^{+28.72}$ & $ 79.56_{-20.41}^{+91.69}$& \textbf{300}&\textbf{300}&- &- &- &-\\
Ionized reflection continuum &  kT $_e$(keV) & - & -& -& -& $2.95\pm0.25$ & $2.72 \pm 0.15$ & $<17.42$& $ <20.36$\\
Ionized reflection continuum &  $\log(n / \rm cm^{-3})$& \textbf{15} &\textbf{15}& $16.81_{-2.42}^{+0.65}$ & $18.01_{-2.73}^{+0.26}$& \textbf{15} & \textbf{15} & $<16.17$ & $<16.86$ \\
Ionized reflection continuum& R $_{\mathrm{refl}}$ & $>8.01$ &  $1.03_{-0.41}^{+0.52}$ &$5.09_{-2.53}^{+2.25}$ &$0.57_{-0.18}^{+0.52}$& $6.19\pm1.15$ & $4.47^{+1.26}_{-1.03}$ & $>6.58$ & $>3.85$ \\
\hline
Goodness of fit &$\frac{\chi^2}{ \rm dof} = \chi^2_r$ &$\frac{854}{788}=1.08$ & $\frac{847}{787}=1.07$ & $\frac{862}{788}=1.09$ & $\frac{844}{787}=1.07$&$\frac{864}{808}=1.06$ &$\frac{854}{808}=1.06$ &$\frac{876}{808}=1.08$ & $\frac{840}{808}=1.04$\\
\hline
\hline
\end{tabular}
\end{table}

\end{landscape}
\end{center}

\bsp	
\label{lastpage}
\end{document}